\documentclass[11pt]{article}
\usepackage{axodraw}
\usepackage{epsfig}
\usepackage{amsfonts}
\newcommand{\la}[1]{\label{#1}}
\newcommand{\be}{\begin{equation}}
\newcommand{\ee}{\end{equation}}
\newcommand{\ba}{\begin{eqnarray}}
\newcommand{\ea}{\end{eqnarray}}
\newcommand{\bi}{\begin{itemize}}
\newcommand{\ei}{\end{itemize}}
\newcommand{\rmi}[1]{{\mbox{\scriptsize #1}}}
\newcommand{\nr}[1]{(\ref{#1})}
\newcommand{\tr}{{\rm Tr\,}}

\newcommand{\Hc}{{\rm H.c.\ }}

\newcommand{\nn}{\nonumber \\}
\newcommand{\fr}[2]{{\frac{#1}{#2}}}
\newcommand{\msbar}{\overline{\mbox{\rm MS}}}

\renewcommand{\vec}[1]{{\bf #1}}

\newcommand{\bmu}{\bar\mu}
\newcommand{\trp}{{\rm Tr}_{\rm v}\,}

\newcommand{\Nf}{N_{\rm f}}
\newcommand{\Nv}{N_{\rm v}}
\newcommand{\Pv}{P_{\rm v}}
\newcommand{\Nc}{N_{\rm c}}
\newcommand{\rmO}{{\rm O}}
\newcommand{\xpt}{$\chi$PT\ } 
\newcommand{\zz}%
 {{\mathbb{Z}}^4}
\renewcommand{\a}{r}    
\renewcommand{\b}{s}    
\renewcommand{\c}{u}    
\renewcommand{\d}{v}    
\renewcommand{\k}{k}    
\newcommand{\ta}{\tilde r}     
\newcommand{\tb}{\tilde s}     
\newcommand{\tc}{\tilde u}     
\newcommand{\td}{\tilde v}     
\newcommand{\1}{d} 
\newcommand{\2}{u} 
\newcommand{\3}{s} 
\newcommand{\4}{c} 
\newcommand{\gvec}[1]{\bar{#1}}
\newcommand{\ga}{\gvec{\a}}
\newcommand{\gb}{\gvec{\b}}
\newcommand{\gc}{\gvec{\c}}
\newcommand{\gd}{\gvec{\d}}

\newcommand{\hvec}[1]{\hat{#1}}
\newcommand{\ha}{\hvec{\a}}
\newcommand{\hb}{\hvec{\b}}
\newcommand{\hc}{\hvec{\c}}
\newcommand{\hd}{\hvec{\d}}

\newcommand{\mS}{\mathcal{S}}
\newcommand{\mP}{\mathcal{P}}
\newcommand{\mL}{\mathcal{L}}
\newcommand{\mW}{\mathcal{W}}
\newcommand{\RR}{{\rm I\kern -.2em  R}}
\newcommand{\eq}{Eq.~}
\newcommand{\eqs}{Eqs.~}
\newcommand{\fig}{Fig.~}
\newcommand{\figs}{Figs.~}
\newcommand{\se}{Sec.~}

\def\lsi{\raise0.3ex\hbox{$<$\kern-0.75em\raise-1.1ex\hbox{$\sim$}}}
\def\gsi{\raise0.3ex\hbox{$>$\kern-0.75em\raise-1.1ex\hbox{$\sim$}}}
\newcommand{\lsim}{\mathop{\lsi}}
\newcommand{\gsim}{\mathop{\gsi}}

\makeatletter \@addtoreset{equation}{section} \makeatother
\renewcommand{\theequation}{\arabic{section}.\arabic{equation}}
\makeatletter
\renewcommand\section{\@startsection {section}{1}{\z@}%
                                   {-5.5ex \@plus -1ex \@minus -.2ex}
                                   {2.3ex \@plus.2ex}%
                                   {\normalfont\large\bfseries}}
\renewcommand\subsection{\@startsection{subsection}{2}{\z@}%
                                     {-3.25ex\@plus -1ex \@minus -.2ex}%
                                     {1.5ex \@plus .2ex}%
                                     {\normalfont\normalsize\bfseries}}
\renewcommand\thesection {\@arabic\c@section}
\renewcommand\thesubsection   {\thesection.\@arabic\c@subsection}
\renewcommand{\@seccntformat}[1]{%
\csname the#1\endcsname.\hspace{1.0em}}
\makeatother

\input pix.sty

\begin{document}

\begin{titlepage}
\begin{flushright}
BI-TP 2006/26 \\
FTUV-06-0720 \\
IFIC/06-18 \\
hep-lat/0607027\\
\end{flushright}
\begin{centering}
\vfill

\mbox{\Large\bf Probing the chiral weak Hamiltonian at finite volumes}

\vspace*{0.8cm}

P.~Hern\'andez$^{\rm a,}$\footnote{pilar.hernandez@ific.uv.es}
and 
M.~Laine$^{\rm b,}$\footnote{laine@physik.uni-bielefeld.de}

\vspace*{0.8cm}

{\em $^{\rm a}$%
Dpto.\ F\'{\i}sica Te\'orica and IFIC, Edificio Institutos Investigaci\'on, \\
Apt.\ 22085, E-46071 Valencia, Spain\\}

\vspace{0.3cm}

{\em $^{\rm b}$%
Faculty of Physics, University of Bielefeld, 
D-33501 Bielefeld, Germany\\}

\vspace*{0.8cm}

{\bf Abstract}
 
\end{centering}
 
\vspace*{0.4cm}

\noindent
Non-leptonic kaon decays are often described through 
an effective chiral weak Hamiltonian, whose couplings 
(``low-energy constants'') encode all non-perturbative QCD physics. 
It has recently been suggested that these low-energy
constants could be determined at finite volumes by matching the 
non-perturbatively measured three-point correlation 
functions between the weak Hamiltonian and two left-handed flavour currents, 
to analytic predictions following from chiral perturbation theory. Here we 
complete the analytic side in
two respects: by inspecting how small (``$\epsilon$-regime'') and intermediate 
or large (``$p$-regime'') quark masses connect to each other, and by 
including in the discussion the two leading $\Delta I = 1/2$ operators. 
We show that the $\epsilon$-regime offers a straightforward strategy for 
disentangling the coefficients of the $\Delta I = 1/2$ operators, 
and that in the $p$-regime finite-volume effects are significant
in these observables once the pseudoscalar mass $M$ and the box length 
$L$ are in the regime $ML \lsim 5.0$.



\vspace*{1cm}
 
\noindent
September 2006

\vfill
 
\end{titlepage}


\section{Introduction}

Understanding why the $\Delta I = 1/2$ amplitudes for non-leptonic 
kaon decays are so much larger than the $\Delta I = 3/2$ amplitudes, 
is a long-standing problem for QCD phenomenology. It has been known 
since the early 70s that the bulk of the enhancement must be due to 
strong interactions at low energies~\cite{mk}. Therefore a reliable 
explanation must eventually be based on systematic non-perturbative 
methods, in particular on lattice QCD~\cite{lat,b}. 

It was realized long ago that instead of computing directly the 
decay amplitudes with lattice QCD, a simpler alternative is to use lattice 
simulations to determine the relevant low-energy constants (LECs) 
of the effective chiral weak Hamiltonian that describes kaon decays \cite{b}, 
and then use chiral perturbation theory to compute the physical 
amplitudes~\cite{b}--\cite{pp}.  
The determination of the LECs can be achieved by matching certain 
observables computed in lattice QCD and in chiral perturbation 
theory  ($\chi$PT), as close as 
possible to the chiral limit. In this respect it is advantageous 
to approach the chiral limit by first 
extrapolating to small or zero 
quark masses, and increase the volume only afterwards. 
This setup corresponds to the so-called 
$\epsilon$-regime of $\chi$PT ~\cite{GL} (see also Ref.~\cite{N}). 
The power-counting rules in this regime~\cite{GL}
guarantee that the contamination from higher order LECs 
is reduced very significantly. In other words, 
the number of LECs that appear at  
the  next-to-leading order (NLO) in the $\epsilon$-regime of $\chi$PT 
is typically much smaller than that at the 
next-to-leading  order in the standard $p$-regime, 
where the infrared cutoff is provided by the pion mass rather than the volume. 

The matching of lattice QCD and the chiral effective theory in the
$\epsilon$-regime has recently been considered in order to extract 
the strong interaction LECs \cite{qcd}--\cite{qcd3}.  
Subsequently, it has been pursued for the
determination of the weak LECs
that we are interested in~\cite{methods,weak,strategy}, as well
as for the study of baryon properties \cite{baryons}. This progress has been
possible thanks to the advent of Ginsparg-Wilson formulations of
lattice fermions~\cite{gw}--\cite{kn}, which possess an exact chiral
symmetry in the limit of vanishing quark masses. Simulations in this
regime are however challenging on the numerical side, and
Refs.~\cite{methods,current} introduced several important technical
advances in order to make them possible.

In Ref.~\cite{strategy}, a strategy based on these methods has been
proposed to reveal the role that the charm quark mass plays in the $\Delta
I=1/2$ rule. In particular, following the suggestion of
Ref.~\cite{methods}, the observables that are considered 
are three-point correlation functions of two left-handed 
flavour currents and the weak operators. The
first step is the matching of these observables, 
to extract the LECs of the weak chiral
effective Hamiltonian, in a theory with a light charm quark, that is in
a four-flavour theory with an exact SU(4) symmetry in the valence
sector. The results of this computation can be found
in Ref.~\cite{prl}. The next step of the strategy is to increase 
the charm quark mass and
monitor the LECs as we move towards a theory with an SU(3) flavour
symmetry~\cite{strategy,largemc}.

In a previous paper~\cite{weak}, we have already
computed the NLO $\epsilon$-regime predictions 
for the correlators of left-handed flavour currents and the
$\Delta I = 3/2$ weak operator, whose coefficient determines 
the kaon mixing parameter $\hat B_K$ in the chiral limit. 
The purpose of the present paper is to extend the results
of Ref.~\cite{weak} in two ways. First of all, we compute
the same observables as before, but also at larger 
quark masses, corresponding to the $p$-regime of 
chiral perturbation theory. The goal is to obtain a better understanding 
of the regions of validity of the $\epsilon$ and $p$-regimes. 
Second, we include the $\Delta I = 1/2$ weak operators in the analysis. 

We find that the $\epsilon$-regime does offer a clean 
way of disentangling the coefficients of the two
leading-order $\Delta I = 1/2$ operators. 

It is well known that the description of quenched simulations, which
still are widely in use today, through a quenched version of chiral
perturbation theory, is rather problematic. In particular the
$p$-regime is strongly affected by quenched ambiguities that increase
significantly the number of LECs \cite{gp}, making it difficult to
identify those that should be closest to the ones in the full
theory. We have studied the effect of these ambiguities also in the
$\epsilon$-regime at NLO, and find that they are significantly less
severe in this case.

In most of our analysis we will concentrate, 
however, on the full physical theory. The
most immediate applications might then follow through the use of mixed
fermion frameworks~\cite{pq}, though progress towards dynamical
Ginsparg-Wilson fermions is also taking place~\cite{fp}.

It should be made clear from the onset that
choosing to consider correlators 
involving left-handed flavour currents in this paper,
is not meant to indicate that they would necessarily be 
the ultimate way for determining the weak LECs. For instance, 
employing the zero-mode wave functions of the massless Dirac operator 
might also lead to a useful probe, even though for
the pion decay constant they seem to be slightly disfavoured 
in comparison with the left-handed flavour currents~\cite{zeromode}. 

Other methods to obtain the weak LECs have also been 
considered in the literature. For lattice approaches without 
an exact chiral invariance see, e.g., 
the recent work in Refs.~\cite{twm}.
For models inspired by the large-$\Nc$ expansion see, e.g., 
Refs.~\cite{bbg,hpr}. 

This paper is organised as follows. 
We formulate the problem in \se\ref{se:formulation}, 
discuss the various regimes of chiral perturbation theory 
in \se\ref{se:xpt}, address the $\Delta I = 3/2$ operators
in \se\ref{se:32}, and the $\Delta I = 1/2$ operators
in \se\ref{se:12}. We conclude in \se\ref{se:conclusions}.

%
\section{Formulation of the problem}
\la{se:formulation}

We start by considering QCD with $4$ flavours.
The quark part of the Euclidean continuum Lagrangian reads 
\be
 {L}_E = \sum_{r = 1}^{4} \bar \psi_r (\gamma_\mu D_\mu + m_r)\psi_r
 \;, 
\ee
where $r$ is a flavour index;
the Dirac matrices $\gamma_\mu$ are assumed normalised 
such that $\gamma_\mu^\dagger = \gamma_\mu$, 
$\{\gamma_\mu,\gamma_\nu\} = 2 \delta_{\mu\nu}$; 
$D_\mu$ is the covariant derivative; $m_r$ is the quark mass; 
colour and spinor indices are assumed contracted; and repeated indices 
are summed over, even when no summation symbol is shown explicitly. 
In the following we will consider the three lightest quarks as  
degenerate in mass, $m_u = m_d = m_s \equiv m$, while the charm 
quark is heavier, $m_c \gg m$. 

After an operator product expansion in the inverse W boson mass, weak 
interactions can be described with the Fermi theory involving four-quark
operators. 
In the CP conserving case of two generations, the effective 
weak Hamiltonian is then~\cite{mk} (for reviews see, e.g.,~\cite{hg,revs}) 
\be
 H_w = 
 2 \sqrt{2} G_F V_{ud} V^*_{us} 
 \biggl\{
 \sum_{\sigma = \pm 1}
 h_w^\sigma 
 \Bigl( [{O_{w}}]^\sigma_{suud} - [{O_{w}}]^\sigma_{sccd}\Bigr) 
 + h_m 
 [{O_{m}}]_{sd} \biggr\}
 + \Hc \;, \la{Hw}
\ee
where $h_{w}^\pm, h_{m}$ are scheme-dependent
dimensionless Wilson coefficients, with leading order 
values $h_{w}^\pm = 1, h_{m}=0$. 
The coefficients $h_{w}^\pm$ are known to two loops 
in perturbation theory \cite{wilson}, while $h_{m}$ remains undetermined. 
In \eq\nr{Hw} we have introduced the notation 
\ba
 [ O_{w} ]^\sigma_{\a\b\c\d} & \equiv & 
 \fr12 
 \Bigl( [ O_{w} ]_{\a\b\c\d} + 
 \sigma [ O_{w} ]_{\a\b\d\c} \Bigr)\;, \la{Owplus} \\[2mm]
 {[ O_{w} ]}_{\a\b\c\d} & \equiv & 
 (\bar\psi_{\a} \gamma_\mu P_{-} \psi_{\c}) 
 (\bar\psi_{\b} \gamma_\mu P_{-} \psi_{\d}) \;,  
 \la{O_QCD} \\[2mm]
 {[ O_{m} ]}_{sd} ~~ & \equiv & 
 (m_c^2 - m_u^2) \{ m_s (\bar\psi_{\3} P_- \psi_{\1}) + 
 m_d (\bar\psi_{\3} P_+ \psi_{\1}) \} \;.
 \la{Lw_QCD_general} \la{O2_QCD}
\ea
Here $r,s,u,v$ are generic flavour indices, 
while $u,d,s,c$ denote the physical flavours.
The chiral projection operators 
$P_\pm$ read $P_\pm \equiv (1\pm\gamma_5)/2$, 
where $\gamma_5 = \gamma_0 \gamma_1 \gamma_2 \gamma_3$.
The colour and spinor indices are assumed to be contracted 
within the parentheses.

In order to match the Hamiltonian
of~\eq\nr{Hw} to the one in the SU(3) chiral theory, 
the first step is to decompose it into irreducible representations of 
the SU(3)$_L\times$SU(3)$_R$ flavour group, 
present at low energies. The weak operators 
are singlets under SU(3)$_R$, and projecting them 
onto irreducible representations
of SU(3)$_L$, the weak Hamiltonian can be rewritten as 
\ba
 {H}_w & = &   
  2 \sqrt{2} G_F V_{ud} V^*_{us} 
 \biggl\{
  h_w^+ [{\hat {O}_w}]^+_{\3\2\2\1} 
  + \fr15 h_w^+ [R_w]^+_{\3\1} - h_w^- [R_w]^-_{\3\1} - \nn
  & - & \fr12 (h_w^+ + h_w^-) [O_w]_{\3\4\4\1} 
  - \fr12 (h_w^+ - h_w^-) [O_w]_{\3\4\1\4} 
  + h_m [O_m]_{\3\1}
 \biggr\} + \Hc \;, 
 \la{Lw_QCD_su3}
\ea
where 
\ba
 [\hat {O}_w]^+_{\3\2\2\1} & \equiv & 
 \frac{1}{2} \Bigl\{ [O_w]_{\3\2\2\1} + [O_w]_{\3\2\1\2} 
 - \fr15 \sum_{k=u,d,s}
 \Bigl( [O_w]_{\3\k\1\k} + [O_w]_{\3\k\k\1}  \Bigr)\Bigr\} \;, 
 \la{preO27} \\
 {[R_w]}_{\3\1}^{\pm} & \equiv & 
 \fr12 \sum_{k=u,d,s} \Bigl( [O_w]_{\3\k\1\k} \pm [O_w]_{\3\k\k\1} \Bigr) \;.
 \la{preO8}
\ea
The first operator in \eq\nr{Lw_QCD_su3} transforms
under the 27-plet of the SU(3)$_L$ subgroup: it is symmetric under 
the interchange of quark or antiquark indices, and traceless. 
The remaining ones, transforming as ${\bf 3^* \otimes 3}$
and being traceless, belong to irreducible representations 
of dimension 8. 

If, as the next step, the charm quark is also integrated out, 
then the operators in \eq\nr{Lw_QCD_su3} go over into the 
standard ones, commonly denoted by $Q_i$, $i = 1,...,6$~\cite{itep,giwi}
(of which five are independent). It is probably safer to 
keep the charm quark in the simulations, though, since integrating 
it out perturbatively is not guaranteed to be a safe procedure. 
Moreover, the quenched three-flavour theory contains spurious 
operators~\cite{gp}.
For these reasons, we prefer to consider the four-flavour theory of
\eq\nr{Lw_QCD_su3} to be the QCD-side
of our problem. 

Now, at large distances, the physics of QCD can be reproduced by 
chiral perturbation theory. For 
a degenerate quark mass matrix, the leading order chiral Lagrangian reads
\ba
 \mathcal{L}_\rmi{$\chi$PT} \!\! & = & \!\! \frac{F^2}{4} \tr 
 \Bigl[ \partial_\mu U \partial_\mu U^{\dagger} \Bigr] 
 - {m \Sigma \over 2} \tr
 \! \Bigl[ e^{i\theta/\Nf} U + U^{\dagger} e^{-i\theta/\Nf}\Bigr] 
 \;,
 \la{XPT} \la{LE}
\ea
where $U \in $ SU($\Nf$), $\Nf \equiv 3$,
and $\theta$ is the vacuum angle. Apart from $\theta$,   
this Lagrangian contains two parameters, 
the pseudoscalar decay constant $F_{}$ and the 
chiral condensate $\Sigma$. 
At the next-to-leading order 
in the momentum expansion, additional operators
appear in the chiral Lagrangian, 
with the associated low-energy constants 
$L_1,L_2,...$~\cite{gl2}.

Obviously the chiral model can be extended to include 
a weak Hamiltonian~\cite{cronin}. We denote the 
chiral analogue of $H_w$ in~\eq\nr{Lw_QCD_su3} by ${\cal H}_w$.
To again define dimensionless coefficients, 
we write ${\cal H}_w$ in the form~\cite{b,lo} 
\be
  {\cal H}_w \equiv  2 \sqrt{2} G_F V_{ud} V^*_{us}
  \biggl\{ 
  \fr53 g_{27} {\cal O}_{27} 
  + 2 g_8 {\cal O}_8
  + 2 g_8'{\cal O}'_8
  \biggr\} + \Hc  \;,
 \la{Lw_XPT}
\ee
where $g_{27}, g_8$ and $g_8'$ are the low-energy 
constants we are interested in. The operators read
\ba
 {\cal O}_{27} & \equiv & 
 {[\; \hat {\! {\cal O}}_{w}]}^+_{\3\2\2\1} = 
 \fr35\Bigl(  
 {[ {\cal O}_{w} ]}_{\3\2\1\2} + 
 \fr23 {[  {\cal O}_{w} ]}_{\3\2\2\1}
 \Bigr)
 \;, \la{formofO} \\   
 {[{\cal O}_{w}]}_{\a\b\c\d} & \equiv &  
 \frac{F^4}{4} 
 \Bigl(\partial_\mu U U^\dagger\Bigr)_{\c\a}
 \Bigl(\partial_\mu U U^\dagger\Bigr)_{\d\b}
 \;, \la{O_XPT} \\
 {\cal O}_{8} & \equiv & 
 {[ {\cal R}_{w} ]}_{\3\1}^+ =  \fr12 \sum_{k=u,d,s}
 {[ {\cal O}_{w} ]}_{\3\k\k\1}
 \;, \la{formofR} \\
 {\cal O}'_{8} & \equiv & 
  \frac{F^2}{2} m {\Sigma}
  \Bigl( e^{i\theta/\Nf}U 
   + U^\dagger e^{-i\theta/\Nf} \Bigr)_{\1\3} 
   \;, \la{formofO8p}
\ea
where we have made use of $\tr[\partial_\mu U U^\dagger] = 0$
to simplify the chiral versions of~\eqs\nr{preO27}, \nr{preO8}.

In the following, we will find it useful to 
generalize the notation somewhat from
the standard SU(3) case introduced above. 
Let $\Nv \equiv 3$ be the number of valence
flavours, and $\Nf$ the number of degenerate sea flavours in the 
chiral Lagrangian. The standard case corresponds to $\Nf = \Nv$, 
but one can also envisage other interesting situations, 
for instance $\Nf = 4$~\cite{strategy}, or $\Nf \to 0$. 
We note that the simplified forms in \eqs\nr{formofO}, \nr{formofR}  
only apply for $\Nf = \Nv$; 
in general, the combinations 
in \eqs\nr{preO27}, \nr{preO8} 
need to be employed (the generalizations of these combinations
to arbitrary $\Nv,\Nf$ are summarised in Appendix A).
In the remainder of this Section we have in mind the 
case $\Nf = \Nv$ but the formulae are written in a way 
which will be useful in Appendix C, where
we analyse the situation $\Nf \neq \Nv$.

The principal strategy now is to construct three-point functions
by correlating $H_w$ with two left-handed flavour currents
on the QCD side, and to match
to predictions from \xpt for the same objects. In QCD, 
the left-handed flavour current can formally be defined as
\be
 J^a_\mu \equiv \bar\psi T^a \gamma_\mu P_- \psi 
 \;, \la{Jamu}
\ee
where $T^a$ is a traceless generator of the valence group SU($\Nv$), and
all colour, flavour, and spinor indices are assumed contracted. 
Note that $J^a_\mu$ defined this way
is formally purely imaginary.\footnote{%
 The convention in \eq\nr{Jamu} differs by a factor $i$ 
 from that in Ref.~\cite{weak}, but agrees with 
 the convention of Refs.~\cite{current,strategy}. We use 
 this ``unphysical'' convention since it removes a number of 
 unnecessary overall minus signs from the \xpt predictions. 
 } 

The two and three-point correlation functions between 
the left-handed currents and the weak operators, averaged over
the spatial volume, now read~\cite{methods}: 
\ba
  \tr[T^aT^b]{C}(x_0) & \equiv & 
 \int \! {\rm d}^3 x\, 
 \Bigl\langle {J}^a_0(x) {J}^b_0(0) \Bigr\rangle
 \;, \la{Cqcd} \\
 {[{C}_\rmi{R}]}^{ab} (x_0,y_0) & \equiv & 
 \int \! {\rm d}^3 x \int \! {\rm d}^3 y\, 
 \Bigl\langle {J}^a_0(x) 
 {O}_\rmi{R} (0) {J}^b_0(y) \Bigr\rangle
 \;,   \la{C1qcd} 
\ea
where the index $R$ refers to the representation. 

On the \xpt side, the operator
corresponding to~\eq\nr{Jamu} becomes, 
at leading order in the momentum expansion, 
\be
 \mathcal{J}^a_\mu = 
 \frac{F^2}{2} \tr \Bigl[ T^a U \partial_\mu U^\dagger \Bigr] 
 \;.
\ee
The two-point correlation function $\mathcal{C}(x_0)$ is defined
(apart from contact terms) by
\be
 \tr [T^a T^b] \, \mathcal{C}(x_0) = 
 \int \! {\rm d}^3 x \,  \Bigl\langle
 \mathcal{J}^a_0(x) \mathcal{J}^b_0(0) \Bigr\rangle
 \;, \la{Cxpt}
\ee
and the three-point correlation function we are interested in, reads
(again apart from contact terms)
\be
 {[\mathcal{C}_\rmi{R}]}^{ab} (x_0,y_0) \equiv  
 \int\! {\rm d}^3x
 \int\! {\rm d}^3y\, \Bigl\langle \mathcal{J}^a_0(x) 
 \mathcal{O}_\rmi{R}(0) \mathcal{J}^b_0(y) 
 \Bigr \rangle 
 \;. \la{C1xpt}
\ee
Our task is to compute the objects in \eqs\nr{Cxpt}, \nr{C1xpt} 
under certain circumstances, to be specified in the next Section.  

%
\section{Regimes of chiral perturbation theory}
\la{se:xpt}

\begin{figure}[t]


\centerline{%
\epsfysize=7.0cm\epsfbox{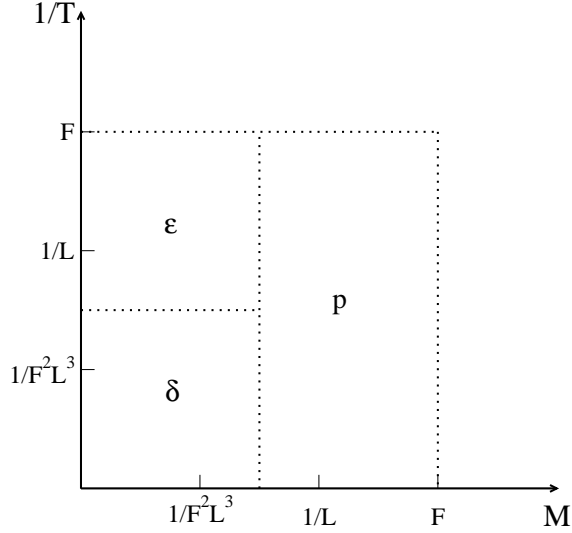}%
}

\caption[a]{\small The different regimes of chiral perturbation
theory, given a fixed spatial extent $L$ of the box, according 
to Ref.~\cite{delta}. Here $T$ is the
temporal extent of the box and $M$ the pseudoscalar mass.
It is assumed that $L \gg 1/F$.} 
\la{fig:regimes}
\end{figure}

Given a fixed spatial extent $L\gg 1/F$ of the box, several different 
kinematical regimes can be identified in $\chi$PT, 
leading to various computational
procedures~\cite{delta}. The situation is summarised
in \fig\ref{fig:regimes}. We will here be interested in the $p$- 
and $\epsilon$-regimes; the $\delta$-regime (corresponding
to small but elongated boxes) is also relevant in principle, 
but quite tedious to handle in practice~\cite{delta}, 
and thus preferably avoided.  

%
\subsection{$p$-regime}

In the $p$-regime, the quark mass is large enough to ensure that
\be
 m \Sigma V \gg 1
 \;. 
\ee
It follows from this condition that the Goldstone field 
$\xi$, defined through $U = \exp(2 i \xi/F)$,    
behaves effectively as a small quantity, and can be expanded in.
Chiral corrections are obtained as an expansion in $\left(M/F\right)^2$ and 
$1/(F L)^2$, where $M^2 \equiv 2 m \Sigma/F^2$.
The power-counting rules in this regime count both of these
expansion parameters at the same order:
\be
 M \sim p 
 \;, \quad
 L \sim \frac{1}{p}
 \;,
\ee
where $p$ is assumed small, $p \ll F$.
The temporal extent $T$ can in principle be small or large, 
as long as $T\gsim 1/p$. Of course, it is also possible to send 
$L\to\infty$ in the $p$-regime expressions. 
The situation is illustrated in \fig\ref{fig:regimes}.

Inserting the Taylor-series 
of $U$ into \eq\nr{LE}, the propagator becomes 
\be
 \Bigl\langle \xi_{\c\a}(x) \, \xi_{\d\b}(y) \Bigr\rangle  = 
 \fr12 \Bigl[\delta_{\c\b} \delta_{\d\a} G(x-y;M^2) - 
 \delta_{\c\a} \delta_{\d\b} E(x-y;M^2) \Bigr]\;, \la{gen_prop}
\ee
where 
\be
 G(x;M^2) = \frac{1}{V} 
 \sum_{n \in \zz} 
 \frac{e^{i p \cdot x}}{p^2+M^2}
 \;, \quad
 p \equiv (p_0,\vec{p}) 
 \equiv 2\pi\Bigl( \frac{n_0}{T}, \frac{\vec{n}}{L} \Bigr)
 \;,
\ee
and $V \equiv T L^3$ is the volume. Here we have also 
set $\theta = 0$, as is usually done in the $p$-regime. 
In the unquenched case, $E(x;M^2) = G(x;M^2)/\Nf$, but 
we keep everywhere $E(x;M^2)$ completely general. The reason is that then the 
form of~\eq\nr{gen_prop} is general enough to contain also the propagator of 
the replica formulation of quenched chiral perturbation theory~\cite{ds,ddhj}. 

For future reference and 
as an example of a NLO result in the $p$-regime, 
we consider
the two-point correlation function in \eq\nr{Cxpt}.
The result can be written as 
\ba
 \mathcal{C}(x_0) & = & \fr{F^2}2 \biggl\{ \biggl[ 
 1 - \frac{\Nf G(0;M^2)}{F^2} + \frac{8 M^2}{F^2} 
 (\Nf L_4 + L_5) \biggr] M^2 P(x_0)
 - \frac{\Nf}{F^2} \frac{{\rm d} G(0;M^2)}{{\rm d} T} \hspace*{1cm}
 \nn & + &  
 \biggl[ 
 \frac{E(0;M^2)}{F^2} - \frac{8 M^2}{F^2} 
 (\Nf L_4 + L_5 - 2 \Nf L_6 - 2 L_8)
 \biggr] M^2 \frac{{\rm d}}{{\rm d} M^2} \Bigl[ M^2 P(x_0) \Bigr]
 \biggr\}
 \;, \la{Ct_p}
\ea
where (for $|x_0| \le T$)
\ba
  P(x_0) & \equiv & \int \! {\rm d}^3 \vec{x} \, G(x;M^2) = 
 \frac{1}{T} \sum_{p_0} \frac{e^{i p_0 x_0 }}{ p_0^2 + M^2 }
 =  \frac{\cosh[M(T/2 - |x_0|)]}{ {2 M} \sinh[MT/2]}
 \;, 
\ea
while
\be
 G(0;M^2) \equiv G_\infty(M^2) + G_V(M^2)
 \;,
\ee
where $G_\infty(M^2)$ is the infinite-volume value,\footnote{%
  The divergence of $G_{\infty}(M^2)$ for $d\approx 4$ cancels 
  against those in the $L_i$'s~\cite{gl2}, 
  cf.\ \eqs\nr{divL4}, \nr{divL6}.}
\be
 G_{\infty}(M^2) \equiv \int \! \frac{{\rm d}^d p}{(2\pi)^d} 
 \frac{1}{p^2 + M^2}
 \;,  
\ee
and the (finite) function $G_V(M^2)$ incorporates all the volume
dependence~\cite{hal},\footnote{%
  In Ref.~\cite{hal} the function $G_V$ was denoted by $g_1$. 
  }
\be
 G_V(M^2) = 
 \frac{1}{(4\pi)^2} 
 \int_0^\infty \frac{{\rm d}\lambda}{\lambda^2} e^{-\lambda M^2}
 \sum_{n \in \zz} 
 \Bigl(1 - \delta^{(4)}_{n,0} \Bigr)
 \exp \Bigl[
 -\frac{1}{4\lambda} \Bigl( 
 T^2 n_0^2 + L^2 
 |\mathbf{n}|^2
 \Bigr) 
 \Bigr]
 \;.
 \la{GV}
\ee
For $M V^{\fr14} \gg 1$, the finite-volume effects
are exponentially small, and we can set $G_V = 0$.

%
\subsection{$\epsilon$-regime}

In the $\epsilon$-regime, the natural dimensionless 
variable is $\mu \equiv m \Sigma V$.
The power counting rules are now
\be
 m\Sigma \sim \epsilon^4
 \;, \quad
 L \sim \frac{1}{\epsilon}
 \;, \quad
 T \sim \frac{1}{\epsilon}
 \;, \la{epsexp}
\ee
where $\epsilon$ is assumed small, $\epsilon\ll F$.
Of course, it is also possible to send $m\to 0$
in the $\epsilon$-regime expressions.
Hence the parameter $\mu$ is 
parametrically of up to order unity.
In this regime, the Goldstone boson zero-mode $U_0$, 
defined by writing $U = \exp(2 i \bar\xi/F)U_0$, where $\bar\xi$
has non-zero momenta only,  
dominates the dynamics, and needs to be treated non-perturbatively.  
Consequently, gauge field topology plays an important role~\cite{ls}, 
and it is useful to give 
the predictions in sectors of a fixed topological charge $\nu$.

As an example, the two-point correlation function
$\mathcal{C}(x_0)$ of \eq\nr{Cxpt} becomes~\cite{h,currents,weak} 
\be
 \mathcal{C}(x_0) 
 = \frac{F^2}{2 T}
 \biggl[
 1 + \frac{\Nf}{F^2}\biggl(
 \frac{\beta_1}{\sqrt{V}} - \frac{T^2 k_{00}}{V} \biggr)
 + \frac{2 T^2 \mu}{F^2 V} \sigma_\nu(\mu) h_1(\hat x_0 ) 
 \biggr] 
 \;, \la{Ct_eps}
\ee
where $\hat x_0 \equiv x_0/T$, 
and the constants $\beta_1$ and $k_{00}$ are related to 
the (dimensionally regularised) value of 
\be
 \bar G(x) \equiv \frac{1}{V} 
 \sum_{n \in \zz }
 \Bigl(1 - \delta^{(4)}_{n,0} \Bigr) \frac{e^{i p\cdot x}}{p^2} 
 \;, 
 \la{Gx}
\ee
by
\be
 \bar G(0) \equiv -\frac{\beta_1}{\sqrt{V}} \;, \quad
 T \frac{{\rm d}}{{\rm d} T} \bar G(0) \equiv \frac{T^2 k_{00}}{V} 
 \;. \la{beta1}
\ee
Introducing $\rho \equiv T/L$ and 
\ba
 \hat \alpha_p(l_0,l_i) & \equiv & 
 \int_0^1 \! {\rm d} t\, 
 t^{p-1} 
 \Bigl[
 S\Bigl( {l_0^2} / {t} \Bigr) 
 S^3\Bigl( {l_i^2} / {t} \Bigr) - 1 
 \Bigr] 
 \;, \la{alphap}
\ea
where $S(x)$ is an elliptic theta-function, 
$S(x) = \sum_{n=-\infty}^{\infty} \exp(-\pi x n^2)
= \vartheta_3(0,\exp(-\pi x))$, 
a numerical evaluation of these coefficients
is possible through (see, e.g., Refs.~\cite{hal,h})
\ba
 \beta_1 & = & 
 \frac{1}{4\pi}
 \Bigl[ 2 - 
 \hat\alpha_{-1}\Bigl( 
 \rho^{\fr34},\rho^{-\fr14}
 \Bigr)
 -
 \hat\alpha_{-1}\Bigl( 
 \rho^{-\fr34},\rho^{\fr14}
 \Bigr)
 \Bigr]
 \;, \\
 k_{00}  & = &  
 \fr1{12} - \fr14\sum_{\vec{n}\neq \vec{0}}
 \frac{1}{\sinh^2(\pi \rho |\vec{n}|)}
 \;.
\ea
Furthermore, 
$\sigma_\nu(\mu) \equiv {\Nf}^{-1}  {\rm d}\{ \ln \det[I_{\nu+j-i}(\mu)] \} / 
{\rm d} \mu$, 
where the determinant is taken over an $\Nf \times \Nf$ matrix, whose 
matrix element $(i,j)$ is the modified 
Bessel function $I_{\nu+j-i}$~\cite{brower,ls}.
The function $h_1(\tau)$ appearing in~\eq\nr{Ct_eps} 
reads (for $|\tau|\le 1$)
\ba
 h_1(\tau) & \equiv & \frac{1}{2}  
 \left[\left(|\tau| - {1 \over 2}\right)^2 - {1 \over 12}\right]
 \;. \la{ph1} 
\ea

%
\subsection{Further remarks}

In the following, we carry out computations according to the 
$p$ and $\epsilon$-countings as outlined above. Other recent
work for related observables has made use of the $p$-regime, 
with $T\gg L$~\cite{sm,bv,cdh}. There have also been extensive
NLO computations at infinite volume~\cite{laso}, which is 
a special limit of the $p$-regime.

Note that if $1/FL \ll 1$ as our power-counting 
rules assume, and we consider an observable 
that is independent of the topological charge $\nu$, 
then the $\epsilon$ and $p$-regimes should in principle 
be continuously connected to each other (cf.\ \fig\ref{fig:regimes}). 
Concretely, for $M L\ll 1$ and $T\sim L$, 
\eq\nr{Ct_p} goes over into~\eq\nr{Ct_eps} with $\mu \gg 1$, 
in which limit the dependence of \eq\nr{Ct_eps} on $\nu$
disappears. Whether such a crossover takes place in practice remains
to be inspected for each observable separately, and gives 
some feeling concerning the convergence of the \xpt computation, 
i.e.,\ whether $1/FL \ll 1$ is satisfied.

%
\section{The $\Delta I = 3/2$ operator}
\la{se:32}

We now address the determination of $g_{27}$, 
considered previously in the $\epsilon$-regime~\cite{weak}.

%
\subsection{$p$-regime}

\begin{figure}[t]
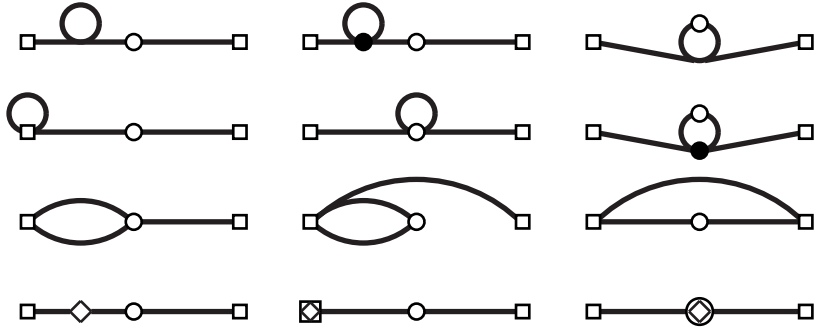


\def\TopoLaction(#1,#2,#3){\piccc{#1(0,15)(16,15) #2(24,15)(40,15)%
#3(40,15)(80,15) %
\SetWidth{1.0} \Line(16,15)(20,19) \Line(20,19)(24,15)%
     \Line(24,15)(20,11) \Line(20,11)(16,15) \SetWidth{1.0}%
\GBoxc(0,15)(5,5){1} \GBoxc(80,15)(5,5){1} \GCirc(40,15){3}{1} }}
\def\TopoLcurrent(#1,#2){\piccc{#1(0,15)(40,15)%
#2(40,15)(80,15) %
\GBoxc(0,15)(7.5,7.5){1} \GBoxc(80,15)(5,5){1} \GCirc(40,15){3}{1}%
\SetWidth{1.0} \Line(-3.5,15)(0,18.5) \Line(0,18.5)(3.5,15)%
     \Line(3.5,15)(0,11.5) \Line(0,11.5)(-3.5,15) \SetWidth{1.0}%
}}
\def\TopoLoperator(#1,#2){\piccc{#1(0,15)(40,15)%
#2(40,15)(80,15) %
\GBoxc(0,15)(5,5){1} \GBoxc(80,15)(5,5){1} \GCirc(40,15){5}{1}%
\SetWidth{1.0} \Line(36,15)(40,19) \Line(40,19)(44,15)%
     \Line(44,15)(40,11) \Line(40,11)(36,15) \SetWidth{1.0}%
 }}

\begin{eqnarray*}
& &
\Topoin(\TLsc,\TLsc,\TLsc,\TAsc) \quad
\Topomassin(\TLsc,\TLsc,\TLsc,\TAsc) \quad
\Topoinop(\TLsc,\TLsc,\TAsc) \\
& & 
\Topocu(\TAsc,\TLsc,\TLsc) \quad
\Topoop(\TAsc,\TLsc,\TLsc) \quad
\Topomassinop(\TLsc,\TLsc,\TAsc) \\
& & 
\Topocuop(\TAsc,\TAsc,\TLsc) \quad
\Topocucuop(\TAsc,\TAsc,\TAsc) \quad
\Topocucu(\TAsc,\TLsc,\TLsc) \\
& & 
\TopoLaction(\TLsc,\TLsc,\TLsc) \quad
\TopoLcurrent(\TLsc,\TLsc) \quad
\TopoLoperator(\TLsc,\TLsc)
\end{eqnarray*}

\caption[a]{\small The NLO
graphs for $\mathcal{C}_{27}$ in the $p$-regime. 
Lines denote meson propagators, 
an open square the left-handed current, 
an open circle the weak operator,
and four-point interactions with no symbol and 
with a closed circle the ``kinetic'' and ``mass'' terms 
in the chiral Lagrangian, respectively. Diamonds indicate
QCD and weak interaction $\rmO(p^4)$ low-energy constants.}
\la{fig:graphs}
\end{figure}

The graphs entering the computation of~\eq\nr{C1xpt} at  
next-to-leading relative order in the $p$-regime are shown 
in~\fig\ref{fig:graphs}, 
with the weak operator $\mathcal{O}_{27}$ to be taken from \eq\nr{formofO}. 
The result can be written in the form 
\be
  {[\mathcal{C}_{27}]}^{ab}(x_0,y_0) = 
   \Delta_{27}^{ab} \, \Bigl[ \mathcal{C}(x_0) \mathcal{C}(y_0) 
 +  \mathcal{D}_{27}(x_0,y_0)  \Bigr]
 \;, \la{c1}
\ee
where (for $\Nv = 3$)
\be
 \Delta_{27}^{ab} = 
 \fr35 T^{\{a}_{ds} T^{b\}}_{uu}
 + \fr25 T^{\{a}_{us} T^{b\}}_{du} 
%
 \;. \la{flavproj}
\ee
As an example, 
choosing kaon and pion type currents, we could take 
\ba
 & &  T^a_{ij} \equiv \delta_{iu} \delta_{js} 
 \,\, \Leftrightarrow \,\, J^a_0 = \bar u \gamma_0 P_- s
 \;, \la{phys1} \\
 & &  T^b_{ij} \equiv \delta_{id} \delta_{ju} 
 \,\, \Leftrightarrow \,\, J^b_0 = \bar d \gamma_0 P_- u
 \;, \la{phys3}
\ea
and then 
\be 
 \Delta_{27}^{ab} = \fr25 
 \;. \la{ix27}
\ee

Given the result
for $\mathcal{C}(x_0)$ in~\eq\nr{Ct_p}, 
the only further missing ingredient
in \eq\nr{c1} is $\mathcal{D}_{27}(x_0,y_0)$. 
We obtain
\ba
 \mathcal{D}_{27}(x_0,y_0) & = & 
 -\frac{F^2}{4} 
 \biggl\{
 \frac{M^2}{T} 
 \frac{{\rm d}^2 G(0;M^2)}{{\rm d} M^2 {\rm d} T}
 + M^2 \frac{{\rm d} G(0;M^2)}{{\rm d} T} 
 \Bigl[
 P(x_0) + P(y_0) 
 \Bigr] 
 + \nn  & & 
 + 2 M^4 G(0;M^2) P(x_0) P(y_0)
 -\fr12 M^4 P(x_0 - y_0)
 \Bigl[
 B(x_0) + B(y_0) 
 \Bigr] 
 + \nn & & 
 + M^4 \int_0^T \! {\rm d}  \tau \, 
 P'(\tau - x_0) P'(\tau - y_0) B(\tau) 
 \biggr\} 
 \;, \la{calD}
\ea
where the new object $B(x_0)$ is defined as (for $|x_0| < T$) 
\ba
 B(x_0) &  = &  \int \! {\rm d}^3 \vec{x} \, \Bigl[ G(x;M^2) \Bigr]^2 = 
 \frac{1}{L^3} \sum_{\vec{p}} 
 \left.
 \biggl[ 
 \frac{\cosh[E(T/2 - |x_0|)]}{ {2 E} \sinh[ET/2]} 
 \biggr]^2 
 \right|_{E \equiv\sqrt{M^2 + \vec{p}^2}}
 \;. \la{Bx0}
\ea

\begin{figure}[t]


\centerline{%
\epsfysize=7.0cm\epsfbox{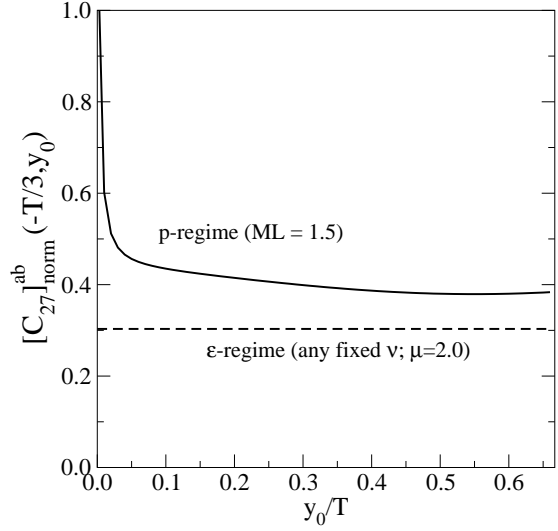}%
}

\caption[a]{\small 
The function ${[\mathcal{C}_{27}]}^{ab}_\rmi{norm}(-T/3,y_0)$. 
The parameters are: $\Nf = 3$, $F = 93$~MeV, 
$L = 2$~fm, $T/L = 2$, 
$\Lambda = 1000$~MeV.
} 
\la{fig:O27y}
\end{figure}

The expression in \eq\nr{calD} is, as such, ultraviolet divergent: 
in dimensional regularization in $d = 4 - 2\epsilon$ dimensions, 
the third and the last terms on the right-hand side
contain poles in $\epsilon$. Denoting $\lambda \equiv -1/32\pi^2\epsilon$, 
we can write 
\be
 \mathcal{D}_{27}(x_0,y_0) =  \mathcal{D}_{27}^r(x_0,y_0)
 + F^2 \lambda
 \Bigl[ 
   \fr12 M^4 P'(x_0) P'(y_0) - M^6 P(x_0) P(y_0) 
 \Bigr]
 \;, \la{divD27}
\ee
where $\mathcal{D}_{27}^r(x_0,y_0)$ is finite. 
The divergences get cancelled against the $\rmO(p^4)$ low-energy constants 
related to weak interactions, as shown in Appendix~B. As there are 
a large number of them, however, it is sufficient for our purposes here
to note that 
the $\rmO(p^4)$ low-energy constants amount to cancelling the 
$1/\epsilon$-divergences in the result and replacing the
corresponding $\msbar$ scheme scale parameter $\bmu$ by two
different physical scales,  
$\Lambda$ for the coefficient of $P(x_0)P(y_0)$
and $\Lambda'$ for the coefficient of $P'(x_0) P'(y_0)$. 

For practical applications, it is convenient to normalise the 
three-point correlator by dividing with two two-point correlators: 
\be
 {[\mathcal{C}_{27}]}^{ab}_\rmi{norm}(x_0,y_0) \equiv
 \frac{{[\mathcal{C}_{27}]}^{ab}(x_0,y_0)}
 {\mathcal{C}(x_0) \mathcal{C}(y_0) } = 
 \Delta_{27}^{ab} \Bigl[1 + 
 \frac{\mathcal{D}_{27}(x_0,y_0)}{\mathcal{C}(x_0) \mathcal{C}(y_0)} \Bigr]
 \equiv
 \Delta_{27}^{ab} \Bigl[1  + 
 \mathcal{R}_{27}(x_0,y_0) \Bigr]
 \;. \la{ratiodef}
\ee
The function $\mathcal{R}_{27}(x_0,y_0)$ is then trivially obtained 
from \eqs\nr{calD} and \nr{Ct_p}; in \eq\nr{Ct_p},
it is even enough to keep the leading order contribution only, 
since $\mathcal{D}_{27}(x_0,y_0)$ gets generated only at NLO.

\begin{figure}[t]


\centerline{%
\epsfysize=7.0cm\epsfbox{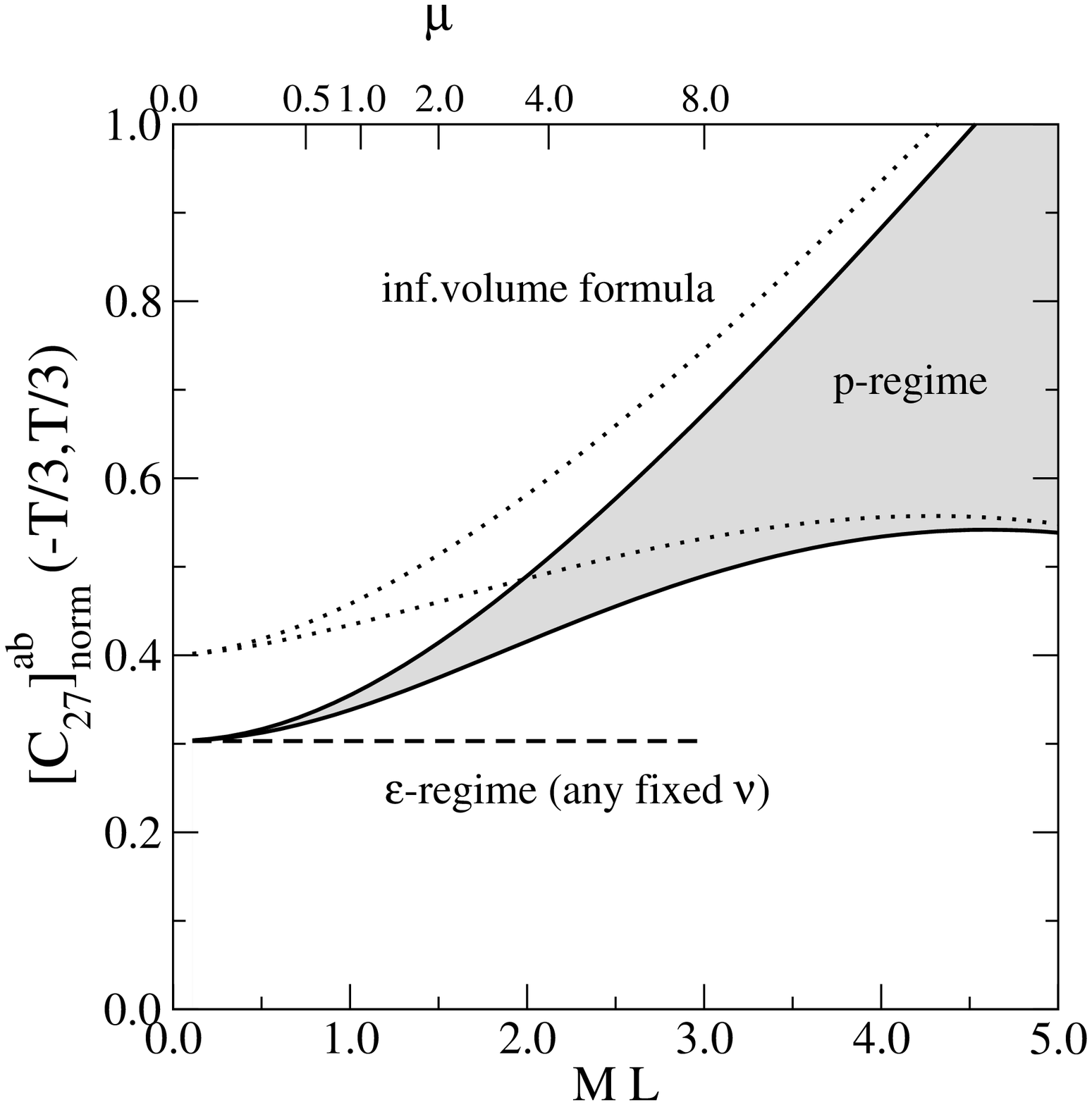}%
~~\epsfysize=7.0cm\epsfbox{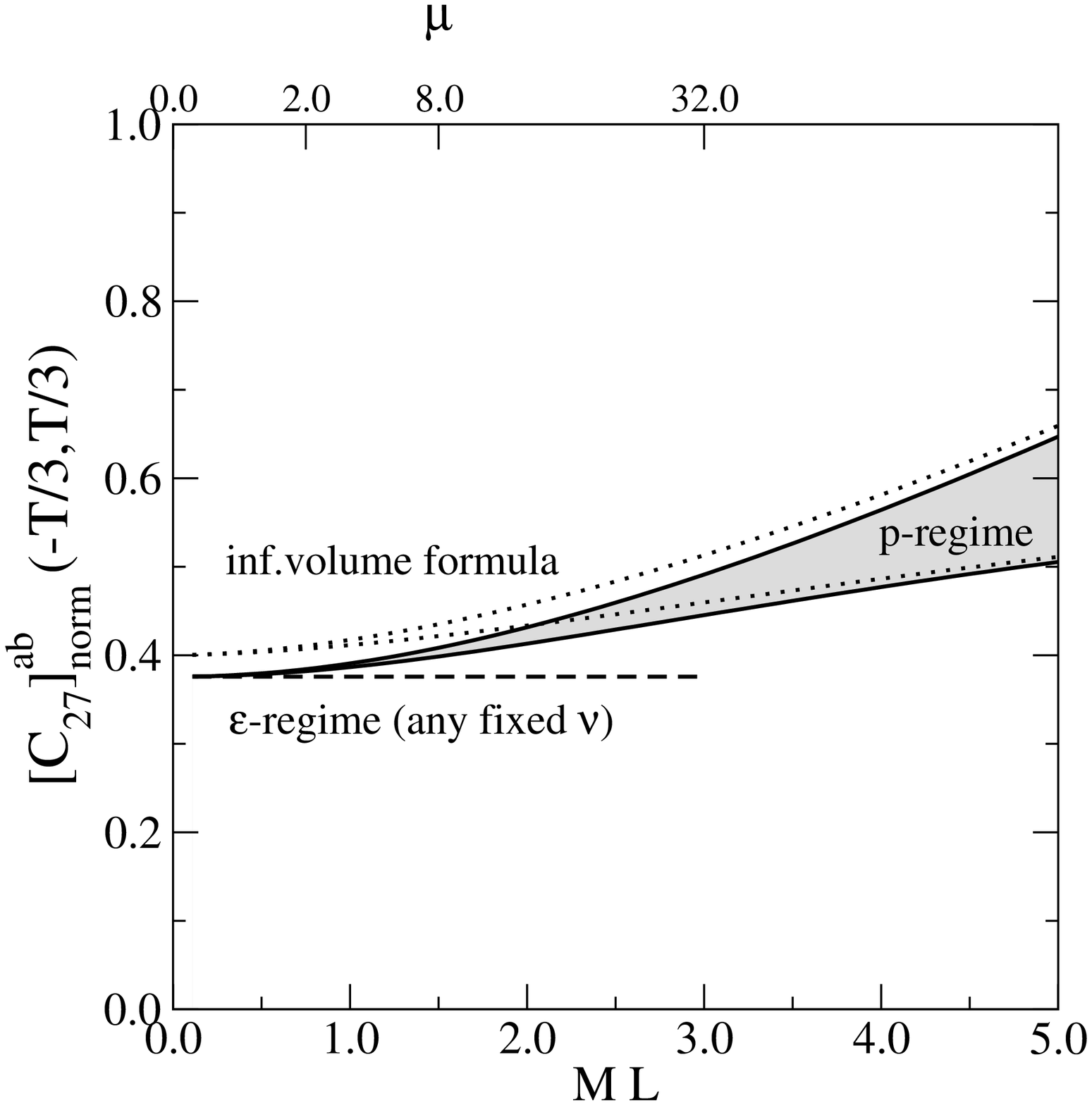}%
}

\caption[a]{\small
The values of 
$[\mathcal{C}_{27}]^{ab}_\rmi{norm}(-T/3,T/3)$.
The parameters are: $\Nf = 3$, $F = 93$~MeV, 
$L = 2$~fm (left), 
$L = 4$~fm (right), 
$T/L = 2$, $\Lambda = (500 - 2000)$~MeV.
} 
\la{fig:O27ML}
\end{figure}

As an example, 
the function ${[\mathcal{C}_{27}]}^{ab}_\rmi{norm}(-T/3,y_0)$ is plotted
in \fig\ref{fig:O27y} as a function of $y_0$, for the index choice in 
\eq\nr{ix27} (solid line). The values of 
${[\mathcal{C}_{27}]}^{ab}_\rmi{norm}(-T/3,T/3)$ are shown in
\fig\ref{fig:O27ML}, as a function of $ML$ 
(the region bounded by solid lines). 
In these plots, the effects of 
the weak LECs have been collected to a single scale 
$\Lambda = \Lambda'$ appearing inside the logarithms, and the scale 
has been varied in a wide range, to indicate the size of 
the uncertainty related to the unknown higher order LECs. 

We would like to stress at this point  
that the $p$-regime results are 
parametrically valid only in the range $ML \gsim 1/FL$: 
for generic observables, the contributions of the Goldstone
zero-modes become dominant if this inequality is not satisfied, 
and need to be resummed, leading to the rules of the $\epsilon$-regime. 
It turns out~\cite{strategy}, however, that in 
the normalised observable ${[\mathcal{C}_{27}]}^{ab}_\rmi{norm}(x_0,y_0)$ 
that we have 
considered here, the contributions from the Goldstone zero-modes 
cancel out at this order. 
Therefore the result can in fact formally be expanded 
as a Taylor-series in $(ML)^2$, with the zeroth order term 
agreeing with the result of the $\epsilon$-regime (see below). Still, one has 
to keep in mind that the Taylor-expanded result only needs to reproduce 
the correct mass dependence in the range  $ML \gsim 1/FL$. 

Let us finally briefly touch the conventional limit of large 
volumes. We assume  $x_0 = -|x_0|, y_0 = |y_0|$, such that 
the charges are on opposite sides of the operator. 
Then $P(x_0) = \exp(-M |x_0|)/2 M$ 
and $P'(x_0)P'(y_0) = - M^2 P(x_0)P(y_0)$.
In other words, 
the distinction  disappears between the two structures getting 
contributions from the higher order LECs (cf. \eq\nr{divD27}), 
just as would happen if a partial integration could be carried out 
with respect to the position of the weak operator. Consequently, 
only a single combination of LECs appears, and the corresponding
effects can be collected into a single scale $\Lambda$.
We obtain 
\be
 \mathcal{R}_{27}(x_0,y_0) = 
 \frac{M^2}{(4\pi F)^2}
 \biggl[ 
 3 \ln\frac{\Lambda^2}{M^2} + 2 - 
 e^{-2 M |x_0|}\phi(2 M |x_0|) -  
 e^{-2 M |y_0|}\phi(2 M |y_0|) 
 \biggr] 
 \;, \la{Rinfvol}
\ee
where 
\ba
 \phi(x) & \equiv & 
 \int_0^\infty \! {\rm d} z \, z^{\fr12} e^{-x z}
 \frac{\sqrt{2+z}}{1+z}
 \biggl[ 
 \frac{1}{2+z} + \frac{1}{1+z} - 2
 \biggr]
 \;. \la{phidef}
\ea
The $x_0$ and $y_0$-dependences in \eq\nr{Rinfvol}
are very small in practice.  
As seen in \fig\ref{fig:O27ML} (dotted line), one needs to 
go to volumes as large as $ML \gsim 5$ in order for the simple 
infinite-volume approximation to be accurate for this 
observable.\footnote{%
 Note that finite-volume corrections depend on the observable
 in question; in particular, the finite-volume effects that we find 
 are much larger than those in typical two-point correlation functions.}
%

%
\subsection{$\epsilon$-regime}

The $\epsilon$-regime results 
for $\mathcal{D}_{27}(x_0,y_0)$
were derived in Ref.~\cite{weak} but, 
for completeness and future reference, 
we briefly reinstate them here. For $\mathcal{D}_{27}$ in \eq\nr{c1}
one obtains 
\ba
 \mathcal{D}_{27}(x_0,y_0) & = & 
 -\frac{F^2}{2 T^2}
 \biggl(1 + T \frac{{\rm d}}{{\rm d} T} \biggr)
 {\bar G}(0)
 \;, \la{calD0} 
\ea
and, using \eq\nr{beta1} as well as the leading-order part
of \eq\nr{Ct_eps}, the ratio in \eq\nr{ratiodef} becomes
\ba
 \mathcal{R}_{27}(x_0,y_0) & = & 
 \frac{2}{(FL)^2}
 \Bigl[ \rho^{-\fr12} {\beta_1}  - \rho\, k_{00} 
 \Bigr]
 \;, \la{calR0}
\ea
where $\rho = T/L$. Note that this result is independent of 
the topological charge $\nu$, 
although computed in a fixed topological sector.

The $\epsilon$-regime prediction for
the function ${[\mathcal{C}_{27}]}^{ab}_\rmi{norm}(-T/3,y_0)$ is plotted
in \fig\ref{fig:O27y} as a function of $y_0$, for the index choice 
in \eq\nr{ix27} (dashed line). The values of 
${[\mathcal{C}_{27}]}^{ab}_\rmi{norm}(-T/3,T/3)$ are shown in
\fig\ref{fig:O27ML}, as a function of $\mu$ (dashed line). 

%
\subsection{Further remarks}
\la{se:rem1}

Let us inspect \fig\ref{fig:O27ML}(left), 
around the region $ML \sim 1.5$, or $\mu \sim 2.0$. Moving 
to smaller values of $\mu$, the $\epsilon$-regime becomes 
more accurate, while at larger $ML$, the $p$-regime should
be the correct procedure. But which result represents better
the truth at this intermediate point, where both countings
are in principle parametrically applicable?

Let us note that 
for the semi-realistic parameters used in \fig\ref{fig:O27ML}(left), 
$1/FL \approx 1.1$.
Therefore, the parametric rules we have assumed are at best satisfied
by a narrow margin. Consequently, higher order corrections in \xpt can 
be important. In the absence of an explicit computation thereof, 
it remains to be inspected phenomenologically
which of the predictions reproduces better the volume and mass 
dependences of the simulation results in this regime.

We end with a small remark on quenching. 
Employing the replica formulation~\cite{ds,ddhj}
of quenched chiral perturbation theory~\cite{BG,S}, 
the only changes with respect to the unquenched situation are
that we need to replace the propagator of~\eq\nr{gen_prop} through
\ba
 E(x;M^2) 
 & \equiv & \frac{\alpha}{2 \Nc} G(x;M^2) + \frac{m_0^2 - \alpha M^2}{2 \Nc} 
 H(x;M^2)
 \;,  \label{qprop}
 \\
 H(x;M^2) & \equiv & \frac{1}{V}
 \sum_{n \in \zz} 
 \frac{e^{i p \cdot x}}{(p^2+M^2)^2}
 \;, 
\ea
where new parameters related to axial singlet field, 
$m_0^2/2 N_c,\alpha/2 N_c$, have been introduced; 
and take $\Nf\to 0$ at the end of the computation. Given that our 
results for ${[\mathcal{C}_{27}]}^{ab}_\rmi{norm}(x_0,y_0)$
are completely independent of $\Nf$ and of the function $E(x;M^2)$, 
however, there is no change with respect to the unquenched 
theory for this observable~\cite{weak}.

%
\section{The $\Delta I = 1/2$ operators}
\la{se:12}

In the case of the $\Delta I = 1/2$ transitions, 
two operators with the right symmetries appear in \eq\nr{Lw_XPT}.
This means that if we have measured some correlation function 
on the QCD side, with an operator $O_8$ transforming 
in the octet representation, then this is to be matched to a linear 
combination of correlation functions on the \xpt side: 
\be
 \int \! {\rm d}^3 x \int \! {\rm d}^3 y\, 
 \Bigl\langle {J}^a_0(x) 
 \, h_8\, {O}_8 (0) \, {J}^b_0(y) \Bigl\rangle \equiv
 g_8 \, {[\mathcal{C}_8]}^{ab}(x_0,y_0) 
 + g_8' \, {[\mathcal{C}_8']}^{ab}(x_0,y_0) 
 \;, \la{match}
\ee
where $h_8$ is the Wilson coefficient, and 
$g_8, g_8'$ are the partial contributions from 
$h_8\, O_8$ to the corresponding LECs. 
We thus have to consider two
different classes of correlators on the \xpt side, in order to be able 
to disentangle the coefficients of these operators. 

%
\subsection{$p$-regime}

For the operator $\mathcal{O}_8$ of \eq\nr{formofR}, 
the graphs entering the computation
of the correlation function in \eq\nr{C1xpt} are the same as 
in~\fig\ref{fig:graphs}, and the correlation function 
has the same form as in \eq\nr{c1}: 
\be
  {[ \mathcal{C}_{8} ]}^{ab}(x_0,y_0) \equiv 
   \Delta_{8}^{ab}\,\Bigl[ \mathcal{C}(x_0)\, \mathcal{C}(y_0) 
 + \mathcal{D}_{8}(x_0,y_0) \Bigr]  
 \;, \la{c8}
\ee
where 
\be
 \Delta_{8}^{ab} = \fr12 \{T^a,T^b \}_{ds} 
 \;, \la{Theta}
\ee
and the function $\mathcal{C}(x_0)$ 
is still given by \eq\nr{Ct_p}.
For the matrices $T^a,T^b$ in \eqs\nr{phys1}, \nr{phys3},
the group theory factor evaluates to 
\be 
 \Delta_{8}^{ab} = \fr12 
 \;. 
\ee
The function $\mathcal{D}_{8}(x_0,y_0)$ reads
\ba
 \mathcal{D}_8(x_0,y_0) & = & -\frac{\Nv}{2} \mathcal{D}_{27}(x_0,y_0) 
 + \nn & + & 
 F^2 M^2 \frac{\Nv + 2}{8} \biggl\{
  \Bigl[ G(0;M^2) - 2 E(0;M^2) \Bigr] P'(x_0) P'(y_0)
 + \nn & & \hspace*{0.0cm} +
   M^2 P(x_0 - y_0)
  \Bigl[
   \tilde B(x_0) + \tilde B(y_0) 
   -\fr12 B(x_0) -\fr12 B(y_0) 
  \Bigr] 
 + \nn & & \hspace*{0.0cm} +
   P'(x_0 - y_0)
   \Bigl[
    \tilde B'(y_0) + \tilde B_0(y_0) + 
    \fr12 B'(x_0) -
    \tilde B'(x_0) - \tilde B_0(x_0) 
    - \fr12 B'(y_0) 
   \Bigr] 
 + \nn & & \hspace*{0.0cm} +
 M^2 \int_0^T \! {\rm d}  \tau \,
 \Bigl[ M^2 B(\tau) - 2 M^2 \tilde B(\tau) - \tilde B_{00}(\tau) \Bigr] 
 P(\tau - x_0) P(\tau - y_0)  
 \biggr\} \;. \la{D8}
\ea
The new objects appearing here are defined as 
\ba
 \tilde B(x_0) & \equiv & \int \! {\rm d}^3 \vec{x} \, G(x;M^2) E(x;M^2)
 \;, \\  
 \tilde B_0(x_0) & \equiv & \int \! {\rm d}^3 \vec{x} \, 
  \Bigl[ \partial_0 G(x;M^2) E(x;M^2) - G(x;M^2) \partial_0 E(x;M^2) \Bigr]
 \;, \\
 \tilde B_{00}(x_0) & \equiv & \int \! {\rm d}^3 \vec{x} \, 
  \Bigl[ \partial_0^2 G(x;M^2) E(x;M^2) - G(x;M^2) \partial_0^2 E(x;M^2) \Bigr]
 \;. 
\ea
We recall that in the unquenched theory, $E(x;M^2) = G(x;M^2)/\Nf$, 
and $\tilde B(x_0)$ thus 
agrees with $B(x_0)/\Nf$ as defined through \eq\nr{Bx0}, 
while $\tilde B_{0}(x_0)$, $\tilde B_{00}(x_0)$ vanish.

\eq\nr{D8} again contains divergences: 
in the unquenched theory, 
\ba
 \mathcal{D}_{8}(x_0,y_0) & = &  \mathcal{D}_{8}^r(x_0,y_0)
 + F^2 \lambda
 \Bigl[ \Bigl( \fr12 - \frac{\Nv + 2}{2 \Nf} \Bigr)
   M^4 P'(x_0) P'(y_0) +
 \nn & & \hphantom{\mathcal{D}_{8}^r(x_0,y_0)
 + F^2 \lambda} 
   + \Bigl( \frac{\Nv-2}{4} + \frac{\Nv + 2}{2\Nf} \Bigr)
   M^6 P(x_0) P(y_0) 
 \Bigr]
 \;, \la{divD8}
\ea
where $\lambda \equiv -1/32\pi^2\epsilon$, and 
$\mathcal{D}_{8}^r(x_0,y_0)$ is finite. The cancellation of 
these divergences against 
the $\rmO(p^4)$ LECs is demonstrated in Appendix~B.

Following \eq\nr{ratiodef}, it is convenient to define
a normalised correlation function by dividing with two current-current
correlators, and we thus obtain 
\be
  {[\mathcal{C}_{8}]}^{ab}_\rmi{norm}(x_0,y_0) \equiv
 \frac{{[ \mathcal{C}_{8} ]}^{ab}(x_0,y_0)}
 {\mathcal{C}(x_0) \; \mathcal{C}(y_0) } = 
 \Delta_8^{ab}\Bigl[ 1 + 
 \mathcal{R}_{8}(x_0,y_0) \Bigr]
 \;. \la{ratio8}
\ee
Again, it is enough to use the leading order forms for the 
functions $\mathcal{C}(x_0)$, $\mathcal{C}(y_0)$ in the 
definition of $\mathcal{R}_{8}(x_0,y_0)$, since 
$\mathcal{D}_{8}(x_0,y_0)$ gets generated only at NLO.

In the infinite-volume limit, the distinction between the various 
types of divergences in \eq\nr{divD8}
disappears, as before. Collecting the corresponding LECs
to a single scale $\Lambda$, we obtain (in the unquenched case)
\ba
 \mathcal{R}_8(x_0,y_0) & = & -\frac{\Nv}{2} \mathcal{R}_{27}(x_0,y_0) 
 + 
 \frac{\Nv + 2}{2}\biggl( 1 - \frac{2}{\Nf} \biggr)
 \frac{M^2}{(4\pi F)^2}
 \biggl[
   2 \ln \frac{\Lambda^2}{M^2} + 1 
 + \nonumber \\[2mm] & & \hspace*{3cm} + 
    e^{-2 M |x_0|}\Xi(2 M |x_0|) + 
    e^{-2 M |y_0|}\Xi(2 M |y_0|)  
 \biggr] \;,
\ea
where $\mathcal{R}_{27}(x_0,y_0)$ is from \eq\nr{Rinfvol}, and
\ba
 \Xi(x) & \equiv & 
 \int_0^\infty \! {\rm d} z \, z^{\fr12} e^{-x z}
 \frac{\sqrt{2+z}}{1+z}
 \biggl[ 
 \frac{1}{2+z} - \frac{1}{1+z} + 2 + 4 z
 \biggr]
 \;. \la{Xidef}
\ea

\begin{figure}[t]


\centerline{%
\epsfysize=7.0cm\epsfbox{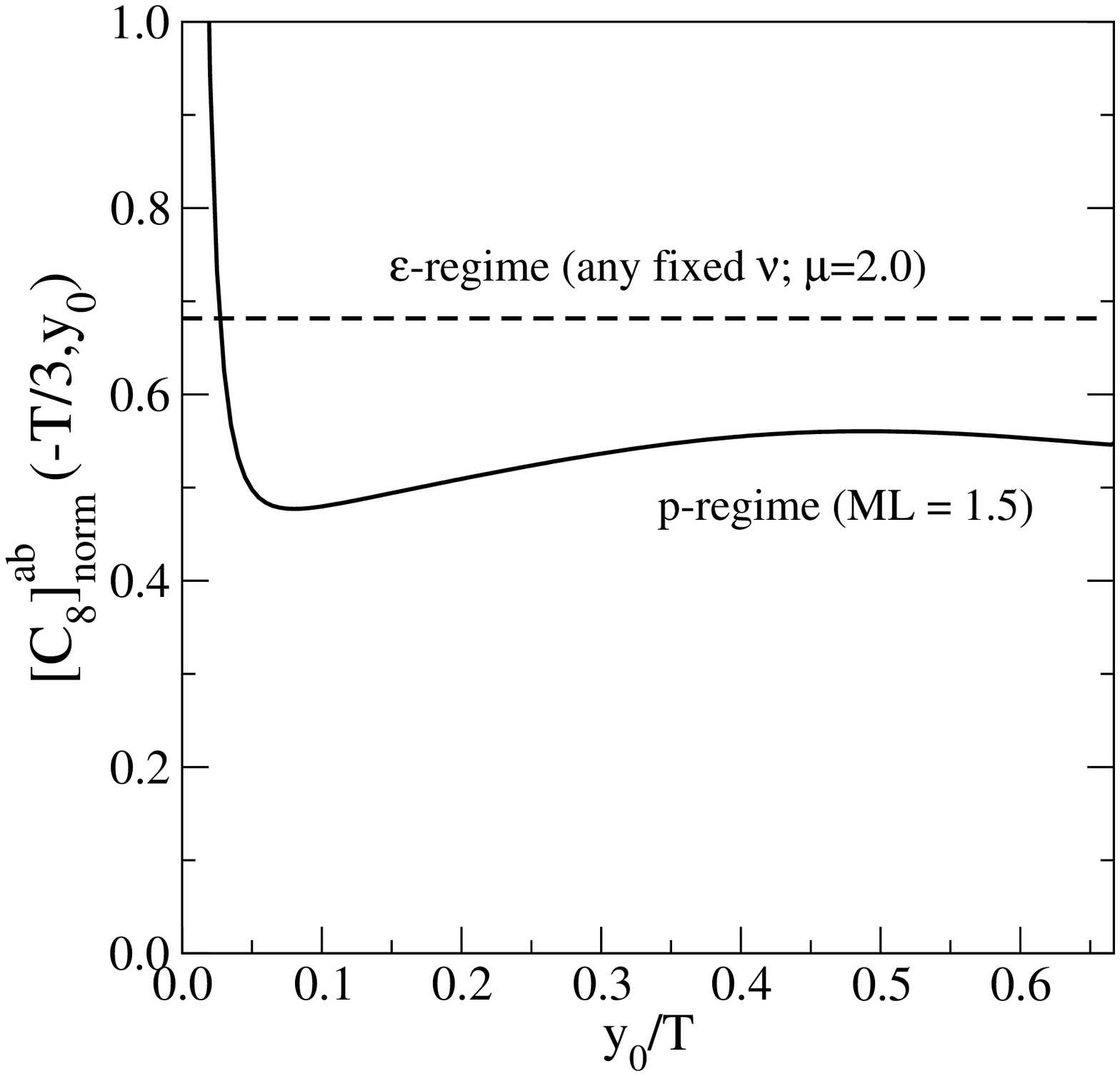}%
~~\epsfysize=7.0cm\epsfbox{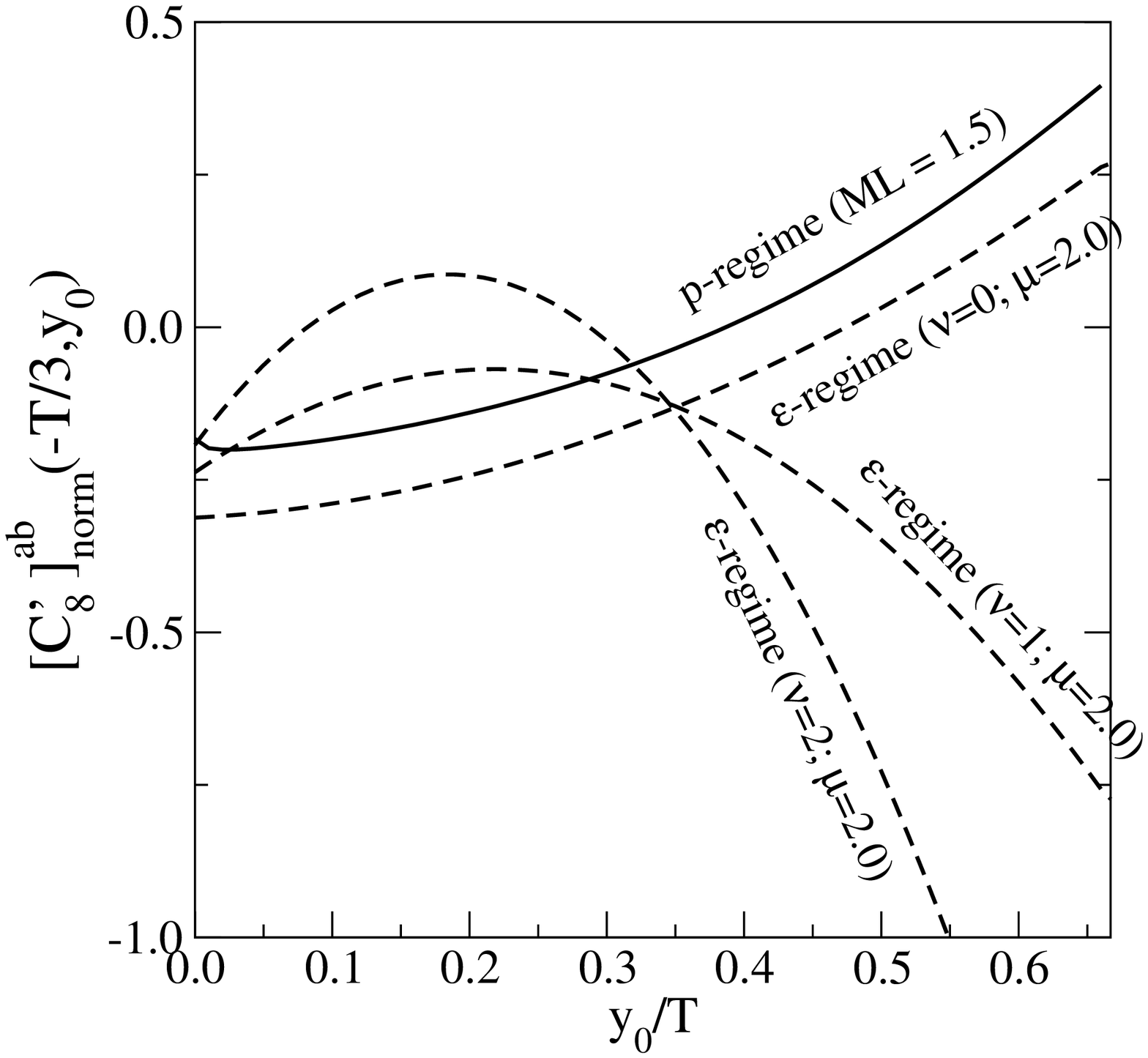}%
}

\caption[a]{\small
Left: the function ${[\mathcal{C}_{8}]}^{ab}_\rmi{norm}(-T/3,y_0)$. 
Right: the function ${[\mathcal{C}_{8}']}^{ab}_\rmi{norm}(-T/3,y_0)$. 
The parameters are: $\Nf = 3$, $F = 93$~MeV, 
$L = 2$~fm, $T/L = 2$, 
$\Lambda = 1000$~MeV.
} 
\la{fig:O8y}
\end{figure}

For the correlator $\mathcal{C}_8'$ the graphs are the same as  
in \fig\ref{fig:graphs} except that, for a vacuum angle $\theta = 0$, 
the weak operator $\mathcal{O}_8'$ only couples to an even number
of Goldstone modes.  The result is now of the form 
\be
 {[\mathcal{C}_{8}']}^{ab}(x_0,y_0) \equiv 
 \Delta_{8}^{ab}\, 
 \mathcal{D}_{8}'(x_0,y_0)
 \;, 
\ee
where
\ba
 \mathcal{D}_{8}'(x_0,y_0) & = & 
 \frac{F^4}{2}\biggl\{ 
 \biggl[ 1 - \frac{\Nf G(0;M^2)}{F^2} + \frac{E(0;M^2)}{F^2} 
 M^2 \frac{{\rm d}}{{\rm d} M^2}
 \biggr] \Bigl[ M^2 P'(x_0) P'(y_0) \Bigr] + 
 \nn & & 
 + \frac{\Nf M^4}{2 F^2}
 \int_0^T \! {\rm d}  \tau \, 
 \Bigl[  P'(\tau - x_0) P'(\tau - y_0) + 
       M^2 P(\tau - x_0) P(\tau - y_0) \Bigr] B(\tau) - 
 \nn & & 
 - \frac{2 M^4}{F^2}
 \int_0^T \! {\rm d}  \tau \, 
 P'(\tau - x_0) P'(\tau - y_0) \tilde B(\tau)
 - \frac{\Nf}{2 F^2}\frac{M^2}{T} 
 \frac{{\rm d} G(0;M^2)}{{\rm d} M^2 {\rm d} T}
 \biggr\} 
 \;. \la{calDp}
\ea
Separating the divergent parts, we get (in the unquenched case)
\ba
 \mathcal{D}_{8}'(x_0,y_0) & = &  \mathcal{D}_{8}'^r(x_0,y_0)
 + F^2 \lambda
 \Bigl[ 
   \Bigl( -\frac{3\Nf}{2} + \frac{3}{\Nf} \Bigr) M^4 P'(x_0) P'(y_0) 
   - \frac{\Nf}{2} M^6 P(x_0) P(y_0)  + 
 \nn 
 & &  \hphantom{\mathcal{D}_{8}^r(x_0,y_0)
 + F^2 \lambda}
 + \frac{1}{\Nf} 
  M^6 \frac{{\rm d}}{{\rm d} M^2} \Bigl( P'(x_0) P'(y_0) \Bigr) 
 \Bigr]
 \;, \la{divD8p}
\ea
where $\mathcal{D}_{8}'^r(x_0,y_0)$ is finite. The cancellation
of divergences is demonstrated in Appendix~B.

\begin{figure}[t]


\centerline{%
\epsfysize=7.0cm\epsfbox{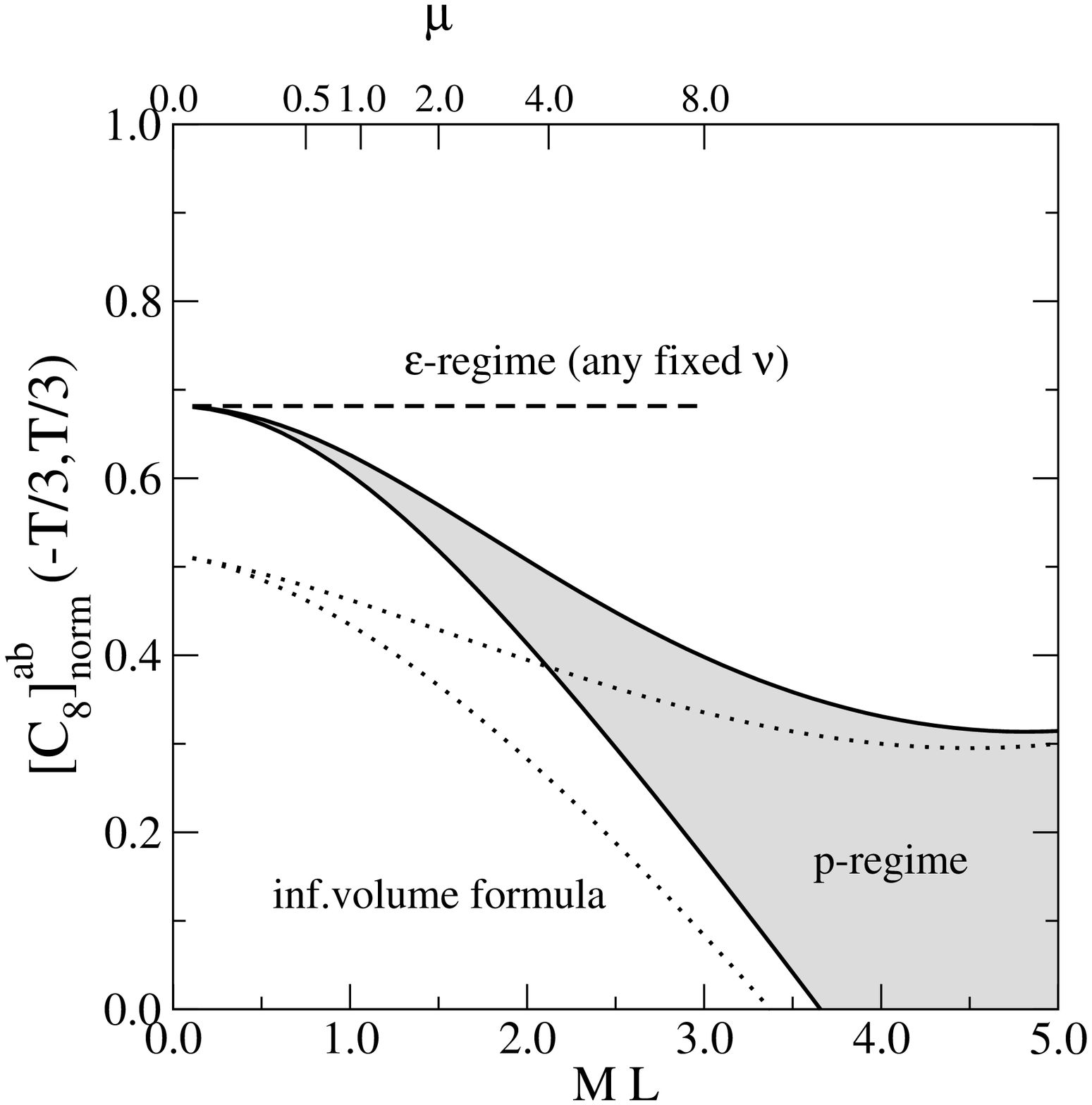}%
~~\epsfysize=7.0cm\epsfbox{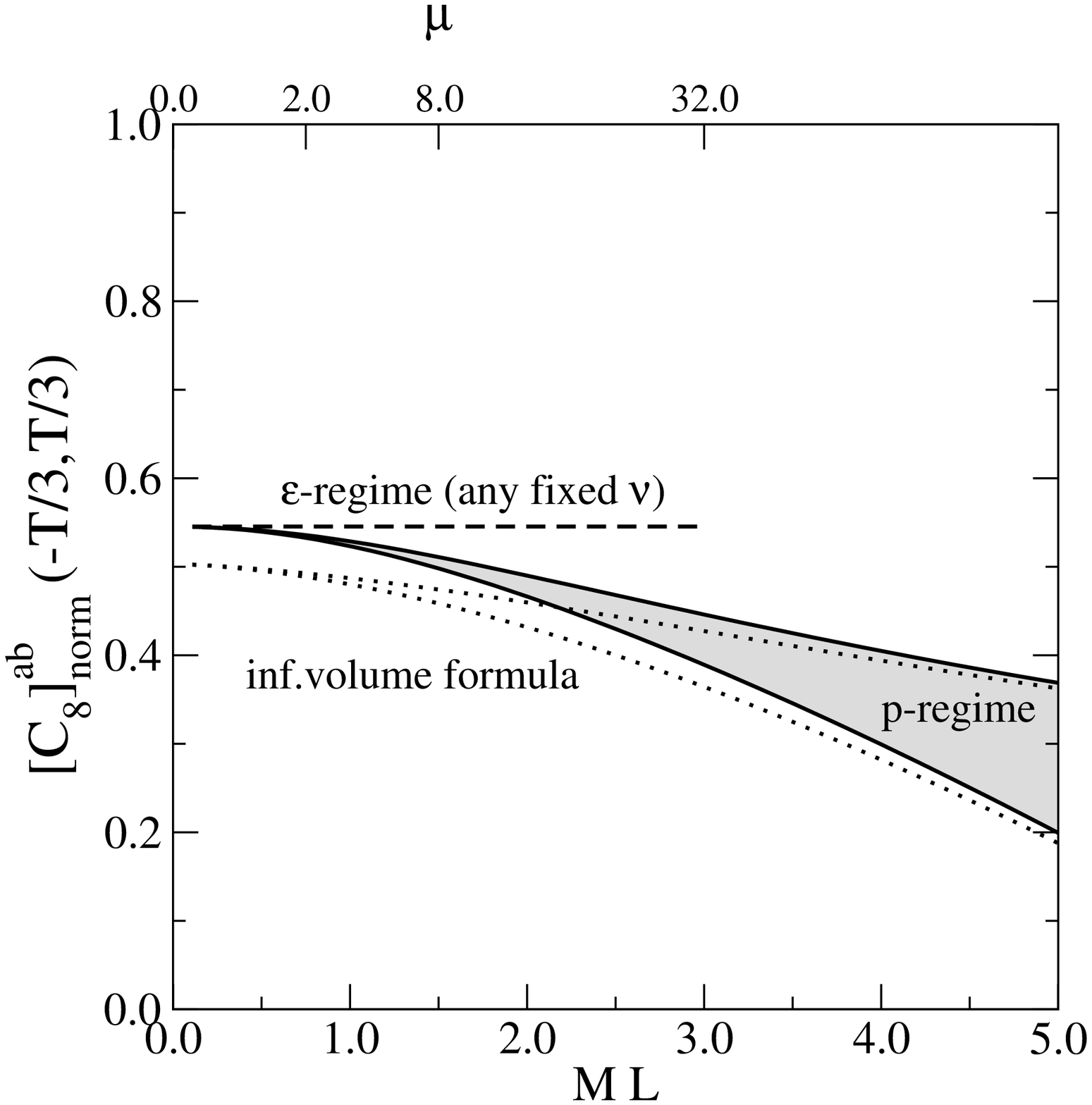}%
}

\caption[a]{\small
The function ${[\mathcal{C}_{8}]}^{ab}_\rmi{norm}(-T/3,T/3)$. 
The parameters are: $\Nf = 3$, $F = 93$~MeV, 
$L = 2$~fm (left), 
$L = 4$~fm (right), 
$T/L = 2$, $\Lambda = (500 - 2000)$~MeV.
} 
\la{fig:O8ML}
\end{figure}

If we want to disentangle the dependences following from 
the operators $\mathcal{O}_8$ and $\mathcal{O}_8'$ in a given 
lattice measurement, we are lead to compare the contributions
from $\mathcal{O}_8'$ with the normalised correlation function 
in~\eq\nr{ratio8}. Therefore, we define 
\be 
  {[\mathcal{C}_{8}']}^{ab}_\rmi{norm}(x_0,y_0) \equiv
 \frac{{[ \mathcal{C}_{8} ']}^{ab}(x_0,y_0)}
 {\mathcal{C}(x_0) \; \mathcal{C}(y_0) }
 \;.
\ee
Treating UV-divergences and higher order LECs as before, 
the correlation functions \linebreak
${[\mathcal{C}_{8}]}^{ab}_\rmi{norm}(-T/3,y_0)$
and
${[\mathcal{C}_{8}']}^{ab}_\rmi{norm}(-T/3,y_0)$ 
are plotted in \fig\ref{fig:O8y}
as a function of $y_0$ (solid lines). The two 
correlators are observed to have a rather different 
dependence on $y_0$, so it is in principle possible to disentangle their 
contributions in a given lattice measurement.
The values of 
${[\mathcal{C}_{8}]}^{ab}_\rmi{norm}(-T/3,T/3)$
and
${[\mathcal{C}_{8}']}^{ab}_\rmi{norm}(-T/3,T/3)$
as a function of $ML$ are illustrated in 
\figs\ref{fig:O8ML}, \ref{fig:O8pML}, respectively
(regions bounded by solid lines).

\begin{figure}[t]


\centerline{%
\epsfysize=7.0cm\epsfbox{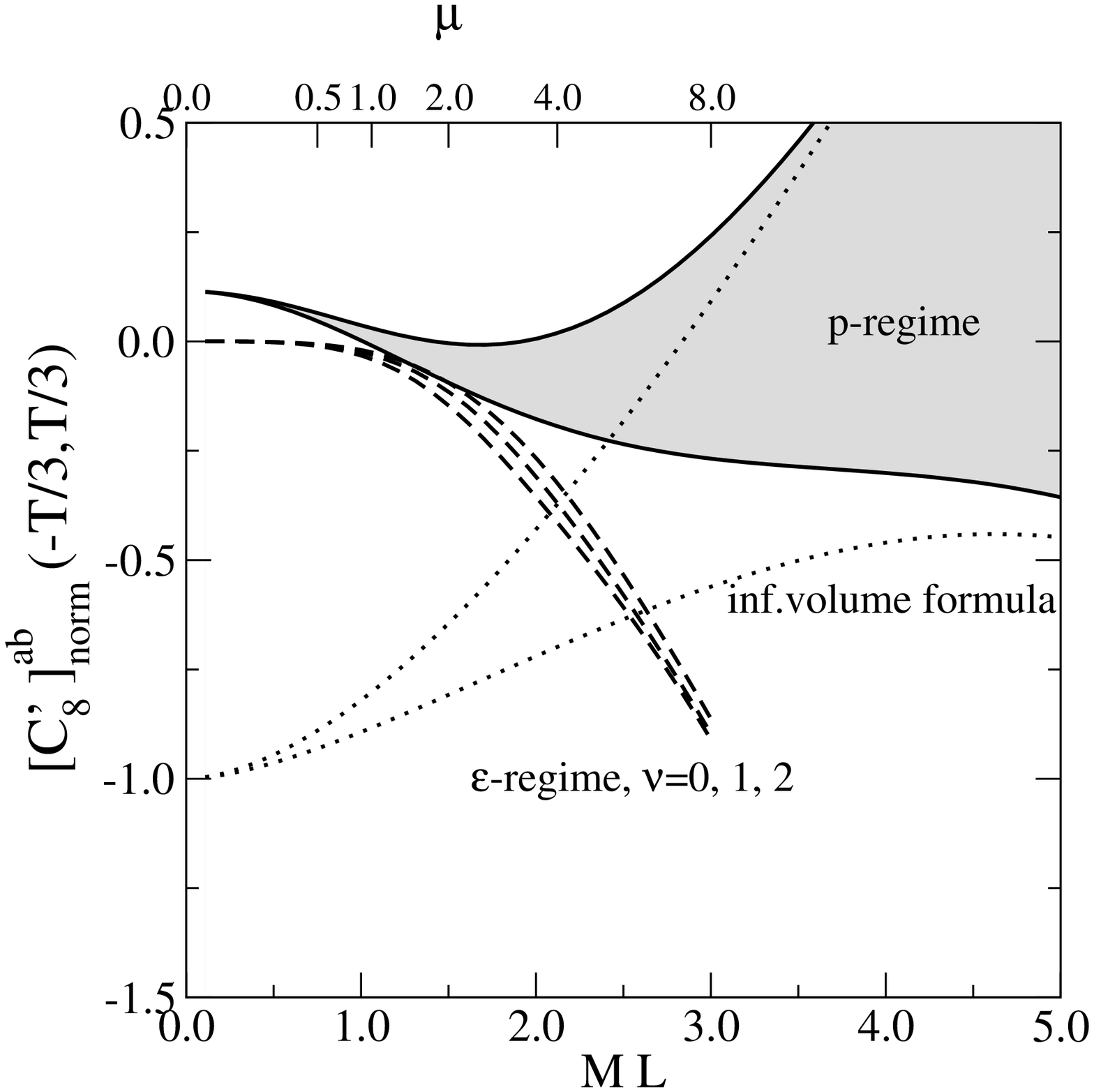}%
~~\epsfysize=7.0cm\epsfbox{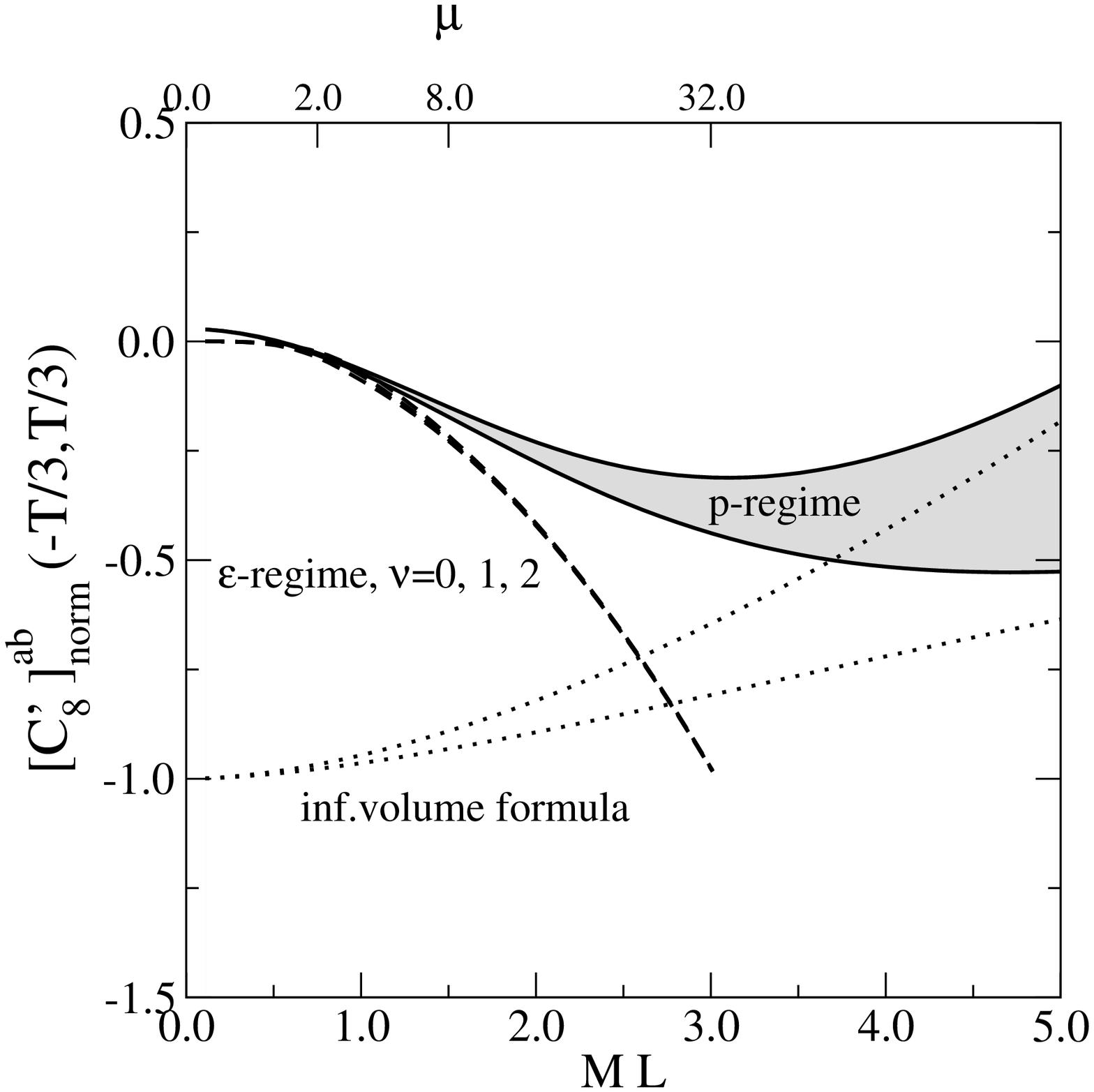}%
}

\caption[a]{\small
The function ${[\mathcal{C}_{8}']}^{ab}_\rmi{norm}(-T/3,T/3)$.
The parameters are: $\Nf = 3$, $F = 93$~MeV, 
$L = 2$~fm (left), 
$L = 4$~fm (right), 
$T/L = 2$, $\Lambda = (500 - 2000)$~MeV.
} 
\la{fig:O8pML}
\end{figure}

It is important to stress that, for $ML \to 0$, the 
correction of relative order $1/F^2$ in \eq\nr{calDp} diverges 
as $\sim 1/F^2M^2 V$. This indicates in a concrete way that 
the $p$-regime computation is no longer reliable for $ML \ll 1/FL$, and we 
need to turn to the $\epsilon$-regime. 
 
Let us again end by commenting on the conventional limit of large 
volumes. Assuming $x_0 = -|x_0|, y_0 = |y_0|$, and inserting the unquenched 
value of $E(x;M^2)$, we obtain for the normalised case 
\ba
 [\mathcal{C}_8']^{ab}_\rmi{norm}(x_0,y_0) & =&  
 - \{ T^a,T^b \}_{ds} 
 \biggl\{ 1 + 
 \frac{M^2}{(4\pi F)^2}
 \biggl[ 
  -\Nf \biggl( 1 + \ln\frac{\Lambda^2}{M^2} \biggr) - 
 \frac{2}{\Nf} \ln \frac{\Lambda^2}{M^2} 
 - 
 \nn & & \hspace*{1cm} 
 - 
 \frac{\Nf}{2} 
 \Bigl\{ 
 e^{-2 M |x_0|}\Delta(2 M |x_0|) +   
 e^{-2 M |y_0|}\Delta(2 M |y_0|) 
 \Bigr\} 
 + 
 \nn & & \hspace*{1cm} 
 + 
 \frac{2}{\Nf} 
 \Bigl\{ 
 e^{-2 M |x_0|}\Upsilon(2 M |x_0|) +   
 e^{-2 M |y_0|}\Upsilon(2 M |y_0|) 
 \Bigr\} 
 \biggr] \biggr\}
 \;, \la{Rpinfvol}
\ea
where 
\ba
 \Delta(x) & \equiv & 
 \int_0^\infty \! {\rm d} z \, z^{\fr12} e^{-x z}
 \frac{\sqrt{2+z}}{1+z}
 \biggl[ 
 \frac{2}{2+z} 
 \biggr]
 \;, \la{deltadef} \\
 \Upsilon(x) & \equiv & 
 \int_0^\infty \! {\rm d} z \, z^{\fr12} e^{-x z}
 \frac{\sqrt{2+z}}{1+z}
 \biggl[ 
 \frac{1}{2+z} + \frac{1}{1+z}
 \biggr]
 \;. \la{gammadef}
\ea
This result is plotted in \fig\ref{fig:O8pML} with dotted 
lines.  We observe again how only values $ML \gsim 5.0$ guarantee 
that finite-volume effects are small 
for our three-point observables.

%
\subsection{$\epsilon$-regime}
\la{ss:o8eps}

\begin{figure}[t]
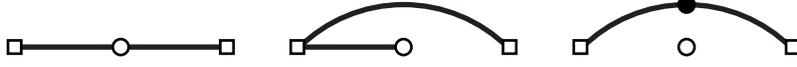


\begin{eqnarray*}
& &
\Topotree(\TLsc,\TLsc) \quad
\Topotreep(\TAsc,\TLsc) \quad 
\Topotreepp(\TAsc) \\
\end{eqnarray*}

\caption[a]{\small The leading-order graphs 
for ${[\mathcal{C}_{8}']}^{ab}(x_0,y_0)$ 
in the $\epsilon$-regime. An open square denotes
the left-handed current,  
an open circle the weak operator, 
and a filled circle a mass insertion.}
\la{fig:C8p}
\end{figure}

We finally move to the $\epsilon$-regime.
For $\mathcal{C}_8$ the graphs are the same as for $\mathcal{C}_{27}$, 
as depicted in Fig.~3 of Ref.~\cite{weak}. The correlator retains the
form in \eq\nr{c8}, with $\Delta_{8}^{ab}$ from \eq\nr{Theta}, 
$\mathcal{C}(x_0)$ from \eq\nr{Ct_eps}, and 
$\mathcal{D}_{27}(x_0,y_0)$, appearing as in \eq\nr{D8}, from \eq\nr{calD0}. 
The order of magnitude of the leading 
term in ${[\mathcal{C}_8]}^{ab}(x_0,y_0)$ is $\rmO(\epsilon^2)$, and  
the NLO term is $\rmO(\epsilon^4)$, while the terms beyond  
$\mathcal{D}_{27}(x_0,y_0)$ in \eq\nr{D8} are formally $\rmO(\epsilon^6)$,
so that the corresponding graph (the sixth in \fig\ref{fig:graphs}) 
can be ignored in the $\epsilon$-regime. Therefore, all information
is in the $\epsilon$-regime version of $\mathcal{D}_{27}(x_0,y_0)$.
To be explicit, 
the normalised form of \eq\nr{ratio8} becomes
\be
 {[\mathcal{C}_{8}]}^{ab}_\rmi{norm}(x_0,y_0) =
 \Delta^{ab}_8\biggl[ 1 -  
 \frac{\Nv}{(FL)^2}
 \Bigl( \rho^{-\fr12} {\beta_1}  - \rho\, k_{00} 
 \Bigr)
 \biggr]
 \;. 
 \la{C8_norm_eps}
\ee
Let us stress, in particular, 
that ${[\mathcal{C}_{8}]}^{ab}_\rmi{norm}(x_0,y_0)$ is independent
of topology and quenching at this order, just like 
${[\mathcal{C}_{27}]}^{ab}_\rmi{norm}(x_0,y_0)$.
(However, as discussed at the end of Appendix~C, quenching does lead 
to the appearance of additional LECs~\cite{gp} that need to be disentangled.)

Let us then address $\mathcal{C}_8'$. Given that 
the $\epsilon$-regime computation is to be carried out at fixed topology, 
the operator $\mathcal{O}_8'$ needs now to be considered in the 
full generality of \eq\nr{formofO8p}, i.e., with a non-vanishing 
vacuum angle $\theta$, unlike in the $p$-regime. Therefore 
$\mathcal{O}_8'$ can also couple to an odd number of Goldstone 
fields. On the other hand, it is easy to see that the tree-level
graphs (cf.\ \fig\ref{fig:C8p}) are already of order $\rmO(\epsilon^4)$.
Comparing with $\mathcal{C}_8$, 
it is therefore enough to restrict to the leading order.
We find
\ba
 {[\mathcal{C}_8']}^{ab}(x_0,y_0) & = & 
 \frac{\mu F^2}{2 V} \biggl\{  \{ T^a.T^b \}_{ds}\;
 \biggl[
   \sigma_\nu(\mu) h_1'(\hat x_0) h_1'(\hat y_0) 
  + 
 \nn & & 
  + 
  \frac{\mu}{1-\Nf^2}
  \biggl\{
  \sigma_\nu'(\mu) 
  + \Nf \sigma_\nu^2(\mu) 
  + \Nf^2 \frac{\sigma_\nu(\mu)}{\mu}
  - \Nf \biggl( 
   1 + \frac{\nu^2}{\mu^2}
   \biggr)
  \biggr\} h_1(\hat x_0 - \hat y_0) \biggr]
  + 
 \nn & &  
  +
 [ T^a,T^b ]_{ds} \frac{\nu}{\mu} 
 \Bigr\{ 
  h_1'(\hat x_0 - \hat y_0) 
  [h_1'(\hat x_0) + h_1'(\hat y_0) ] + 
  h_1(\hat y_0) - h_1(\hat x_0) 
 \Bigr\}  
 \biggr\}
 \;, \hspace*{0.6cm} \la{C8p}
\ea
where $h_1$ is from \eq\nr{ph1}.
The corresponding normalised form reads
\ba
 {[\mathcal{C}_8']}^{ab}_\rmi{norm}(x_0,y_0) & = & 
 \frac{4 T^2}{F^4}
  {[\mathcal{C}_8']}^{ab}(x_0,y_0)
 \;. \la{C8pnorm} \hspace*{1cm}
\ea
This result is to be used in combination with \eq\nr{C8_norm_eps}, 
in order to disentangle the two terms on the right-hand 
side of \eq\nr{match}.

Unlike \eq\nr{C8_norm_eps}, 
the expressions in \eqs\nr{C8p}, \nr{C8pnorm} 
get modified in the quenched theory, 
because they contain Goldstone zero-mode integrals. 
Proceeding as in Ref.~\cite{zeromode}, we find 
\ba
 {[\mathcal{C}_8']}^{ab}_\rmi{q}(x_0,y_0) & = & 
 \frac{\mu F^2}{2 V} \biggl\{  \{ T^a.T^b \}_{ds}\;
 \Bigl[
   \sigma_{\rmi{q}\nu}(\mu) h_1'(\hat x_0) h_1'(\hat y_0) 
  + 
  \mu
  \sigma_{\rmi{q}\nu}'(\mu) 
  h_1(\hat x_0 - \hat y_0) \Bigr]
  + 
 \nn & &  
  +
 [ T^a,T^b ]_{ds} \frac{\nu}{\mu} 
 \Bigr\{ 
  h_1'(\hat x_0 - \hat y_0) 
  [h_1'(\hat x_0) + h_1'(\hat y_0) ] + 
  h_1(\hat y_0) - h_1(\hat x_0) 
 \Bigr\}  
 \biggr\}
 \;, \hspace*{0.6cm} \la{qC8p}
\ea
where the subscript q refers to the quenched theory, and~\cite{dotv}
\be
 \sigma_{\rmi{q}\nu}(\mu) \equiv 
 \mu \Bigl[ I_\nu (\mu) K_\nu (\mu) + I_{\nu +1} (\mu)
 K_{\nu -1}(\mu) \Bigr] +{\nu \over \mu}
 \;, 
 \la{zerocon} 
\ee
where $I_\nu, K_\nu$ are modified Bessel functions. 
Note that \eq\nr{qC8p} could also be obtained from \eq\nr{C8p}
by just naively setting $\Nf\to 0$ and replacing 
$\sigma_{\nu} \to \sigma_{\rmi{q}\nu}$.

Since the functions
${[\mathcal{C}_8]}^{ab}(x_0,y_0)$ and ${[\mathcal{C}_8']}^{ab}(x_0,y_0)$
are not identical, a precise measurement of the time-dependence
of the correlation
function of the left-hand side of \eq\nr{match} would in principle 
make it possible to disentangle the contributions to $g_8, g_8'$. 
In particular, as shown in~\fig\ref{fig:O8y}, any dependence of 
the correlation functions on $\nu$ arises at this order
through the operator $\mathcal{O}_8'$.
In practice, however, the problem emerges that it may not be easy
to obtain such a high accuracy that the two LECs could reliably be
determined from a single observable. Therefore, it may be beneficial to 
define another probe as well, such that the LECs can be disentangled
with better confidence.  
We now show how this can be done. 

%
\subsection{Direct determination of $g_8'$}
\la{ss:direct}

\begin{figure}[t]
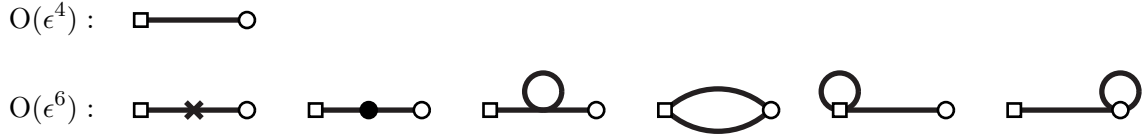


\begin{eqnarray*}
\rmO(\epsilon^4): 
& & \!\!\!
\KTopotree(\TLsc)
\\ 
\rmO(\epsilon^6): 
& & \!\!\!
\KTopomeas(\TLsc,\TLsc) 
\KTopomass(\TLsc,\TLsc) 
\KTopoin(\TLsc,\TLsc,\TAsc) 
\KTopocucu(\TAsc,\TAsc) 
\KTopocu(\TAsc,\TLsc) 
\KTopoop(\TAsc,\TLsc) 
\end{eqnarray*}

\caption[a]{\small The graphs contributing to 
${[\mathcal{K}_{8}']}^a(x_0)$. The notation is as in Fig.~\ref{fig:C8p}, 
with additionally a cross denoting a ``measure term''~(cf.~Ref.~\cite{GL}).}
\la{fig:K8p}
\end{figure}

In order to determine $g_8'$, we consider the correlator 
\be
 {[K_\rmi{R}]}^a (x_0) \equiv
 \int \! {\rm d}^3 x\, 
 \Bigl\langle {J}^a_0(x) 
 {O}_\rmi{R} (0) \Bigr\rangle 
 \;, 
\ee
on the side of QCD, 
and correspondingly 
\be
 {[\mathcal{K}_\rmi{R}]}^a (x_0) \equiv
 \int \! {\rm d}^3 x\, 
 \Bigl\langle \mathcal{J}^a_0(x) 
 \mathcal{O}_\rmi{R} (0) \Bigr\rangle 
\ee
on the \xpt side.
Note that this correlation function is not available in the conventional 
$p$-regime setup (i.e. with $\theta = 0$), 
because it is odd in charge conjugation.

We have computed both ${[ \mathcal{K}_{8} ]}^a(x_0)$ 
and ${[ \mathcal{K}_{8}' ]}^a(x_0)$ at NLO in the $\epsilon$-regime.
Parametrically, the orders of magnitude of the LO and the NLO
graphs are $\rmO(\epsilon^4)$ and $\rmO(\epsilon^6)$, 
respectively. We find, however, that 
at this order ${[ \mathcal{K}_{8} ]}^a(x_0)$ vanishes exactly, 
like in the $p$-regime.

On the other hand, ${[ \mathcal{K}_{8}' ]}^a(x_0)$ does not vanish.
The graphs are shown in \fig\ref{fig:K8p}. 
We find 
\be
 {[ \mathcal{K}_8' ]}^a(x_0)  = - T^a_{ds} \frac{\nu F^2}{V}
 \biggl\{ h_1'(\hat x_0)
 \biggl[
 1 + \biggl( \frac{1}{\Nf} -  \Nf
 \biggr) \frac{\bar G(0)}{F^2} 
 \biggr] 
 + h_2'(\hat x_0)  
 \frac{2 T^2}{F^2 V} 
 \Bigl[ \mu \sigma_\nu(\mu) + \frac{1}{\Nf} 
 \Bigr]
 \biggr\} 
 \;, \la{K8p} \hspace*{1cm}
\ee 
where $\bar G(x)$ is from \eq\nr{Gx}, and (for $|\tau| \le 1$)
\ba
  h_2(\tau) & \equiv & \frac{1}{24} 
 \left[\tau^2 \left(|\tau| - 1\right)^2 - {1 \over 30}\right] 
 \;. \la{h2} 
\ea
The result is illustrated in \fig\ref{fig:K8pplot}, after 
normalisation through 
\be
 \frac{L^3 {[ \mathcal{K}_8' ]}^a(x_0)}{\mathcal{C}(x_0)}
 \equiv
 T^a_{ds} 
 {[ \mathcal{K}_8' ]}_\rmi{norm} (x_0) 
 \;.
\ee

\begin{figure}[t]


\centerline{%
\epsfysize=7.0cm \epsfbox{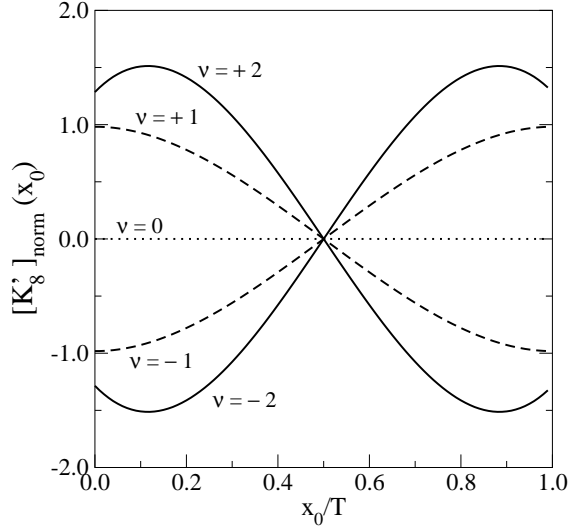}%
}

\caption[a]{\small
The function ${[\mathcal{K}_{8}']}_\rmi{norm}(x_0)$
as a function of $x_0$ and $\nu$.
The other parameters are: $\Nf = 3$, $F = 93$~MeV, 
$\mu = 2.0$, $L = 2$~fm, $T/L = 2$.
} 
\la{fig:K8pplot}
\end{figure}

Repeating the same steps in the quenched theory, we find
\ba
 {[ \mathcal{K}_8' ]}^a_\rmi{q}(x_0)  & = & - T^a_{ds} \frac{\nu F^2}{V}
 \biggl\{ h_1'(\hat x_0)
 \biggl[
 1 + \frac{\alpha}{2\Nc} 
 \frac{\bar G(0)}{F^2} + 
 \frac{m_0^2}{2\Nc} \frac{\bar H(0)}{F^2} 
 \biggr] 
 + 
 \nn & & \hspace*{1.7cm}
 + \frac{2 T^2}{F^2 V} \biggl[ 
 \Bigl( \mu \sigma_{\rmi{q}\nu}(\mu) + \frac{\alpha}{2\Nc} 
 \Bigr)  h_2'(\hat x_0)  
 - \frac{m_0^2 T^2 }{2\Nc}
 h_3'(\hat x_0)
 \biggr\} 
 \;, \la{qK8p} \hspace*{1cm}
\ea 
where $\bar H(x)$ and $h_3(\tau)$ (for $|\tau| \le 1$)
are defined through
\ba
 \bar H(x) & \equiv & \frac{1}{V} 
 \sum_{n \in \zz }
 \Bigl(1 - \delta^{(4)}_{n,0} \Bigr) \frac{e^{i p\cdot x}}{(p^2)^2} 
 \;, \\
 h_3(\tau) & \equiv & \frac{1}{720} 
 \left[ \tau^2 \left(|\tau| - 1\right)^2 
 \left(\tau^2- |\tau| - {1 \over 2}\right) + {1 \over 42}\right] 
 \;. \la{h3} 
\ea
The value of $\bar H(0)$ is given by~\cite{hal}
\ba
 {\bar H}(0) & = & 
 \beta_2 + \frac{\mu^{-2\epsilon}}{(4\pi)^2}
 \biggl[
  \frac{1}{\epsilon} + \ln\Bigl( {\bmu^2} V^{1/2} \Bigr) + 1 + 
  \rmO(\epsilon)
 \biggr] 
 \;, \la{barH}
\ea
where $\ln\bmu^2 \equiv \ln\mu^2 + \ln 4\pi - \gamma_E$, and 
(with $\hat\alpha$ from \eq\nr{alphap})
\ba
 \beta_2 & = & 
 \frac{1}{(4\pi)^2}
 \Bigl[  
 \hat \alpha_0 \Bigl( 
 \rho^{\fr34},\rho^{-\fr14}
 \Bigr) 
 +  
 \hat\alpha_{-2}\Bigl( 
 \rho^{-\fr34},\rho^{\fr14}
 \Bigr) 
 - \fr32 -\ln(4\pi) + \gamma_E
 \Bigr]
 \;. 
\ea
The UV-divergence in \eq\nr{barH} is cancelled by 
$\Sigma$ (cf.\ \eq\nr{formofO8p}), which is to be 
treated as a bare parameter in the quenched theory~\cite{cp}.

To summarise, we now have a method to disentangle the two contributions
related to the LECs $g_8, g_8'$: 
by considering ${[K_8']}_\rmi{norm}(x_0)$, we can 
first match for $g_8'$. Then the corresponding term can be subtracted
from the right-hand side of \eq\nr{match}, and we are able to determine $g_8$.
As illustrated in \fig\ref{fig:O8y}, a cross-check is that the 
dependence on $\nu$ should have disappeared. 

%
\subsection{Further remarks}

The remarks that can be made on the convergence of the 
$\epsilon$ and $p$-regime computations of $\mathcal{C}_8$ 
and $\mathcal{C}_8'$ are largely the same as for 
$\mathcal{C}_{27}$ in \se\ref{se:rem1}. Indeed, 
for $1/FL \ll 1$, there could be a non-vanishing 
overlap, i.e. a regime where both the $p$-regime and 
the $\epsilon$-regime expressions are valid. For 
the more realistic case $1/FL \sim 1$, on the other
hand, this is unlikely to happen. It would be tempting 
to read from \fig\ref{fig:O8pML} that the $p$-regime
expression works in the range $ML \gsim 2.0$, and 
the $\epsilon$-regime expression in the range $\mu \lsim 2.0$, 
but whether this is really the case remains to be seen once
a comparison with lattice simulation results is available.

Concerning quenching, let us stress that 
the correlation 
function ${[\mathcal{C}_8]}^{ab}_\rmi{norm}(x_0,y_0)$ 
is determined in the $\epsilon$-regime
by the same function $\mathcal{R}_{27}(x_0,y_0)$ as 
${[\mathcal{C}_{27}]}^{ab}_\rmi{norm}(x_0,y_0)$, and 
is thus insensitive to quenching at the 
present order. At the same time, the correlation functions 
${[\mathcal{C}_8']}^{ab}_\rmi{norm}(x_0,y_0)$ and
${[\mathcal{K}_8']}^{a}_\rmi{norm}(x_0)$ do get modified.

An important point however is the different relevance of the quenched
ambiguities of Ref.~\cite{gp} in the two regimes.\footnote{%
 We refer here to the ambiguities at the level of the chiral Lagrangian. 
 We assume always that the weak effective Hamiltonian at the quark level 
 contains an active charm so that no ``unphysical'' operators appear in 
 the Operator Product Expansion at the order in the Fermi constant
 at which we are working.} 
In general the
quenched theory contains spurious operators with new LECs. Some of
these originate from the fact that $\Nf \neq \Nv$, a case that is
considered in detail in Appendix~C, while others are related to the
couplings of the axial singlet field that cannot be integrated out in
the quenched limit. The latter modify the terms that in the full
theory would be divergent in the limit $\Nf \rightarrow 0$. We indeed
confirm a rather messy situation in the $p$-regime, where many new
couplings enter; thus we have not carried out a systematic study of all
quenching effects in our observables in this regime. On the other
hand, the quenching ambiguities are reduced to a minimum  
at the NLO in the $\epsilon$-regime, with apparently
only one spurious octet LEC contributing to 
${[\mathcal{C}_8]}^{ab}_\rmi{norm}(x_0,y_0)$.  
Therefore certain octet couplings {\em can} 
be determined by matching the lattice simulation results 
to ${[\mathcal{C}_8]}^{ab}_\rmi{norm}(x_0,y_0)$. 
We elaborate on this issue in more detail in Appendix~C, 
particularly around \eq\nr{O8quench}.

%
\section{Conclusions}
\la{se:conclusions}

We have addressed in this paper the determination of
the $\rmO(p^2)$ LECs of the chiral weak Hamiltonian. As probes we 
have used the three-point correlation functions between 
the weak operators and left-handed flavour currents.  
We have computed the three-point correlation functions
up to next-to-leading order in chiral perturbation theory, 
both in the $\epsilon$ and in the $p$-regimes, for all 
three operators that appear in the SU(3) chiral weak
Hamiltonian. 

While the determination of the LEC $g_{27}$, 
which fixes the $\Delta I = 3/2$ amplitude
of the weak decays $K\to \pi\pi$ as well as the kaon mixing
parameter $\hat B_K$ in the chiral limit, appears straightforward, 
the determination of the LECs fixing the $\Delta I = 1/2$ amplitudes
is more demanding in several respects. Even restricting to the idealised
case of full QCD at large volumes, there are two operators 
with the same flavour symmetry, while only the coefficient of
one of them, $g_8$, contributes to the physical 
kaon decays~\cite{b,rc}. Therefore it is important
to come up with a setup which makes it possible to remove the 
contamination from the other operator in a lattice measurement 
of the type that we have considered. 

We have shown here that this challenge can be met 
by going to the $\epsilon$-regime. The two operators contribute
in very different ways to a given three-point correlation function, 
one leading to a topology-dependent and the other to a topology-independent
result. Moreover, we have found a two-point correlator that is only sensitive
to the ``unphysical'' LEC and can be used to fix it. Therefore, it seems
possible in principle to disentangle the physical 
coefficient $g_8$ from lattice measurements in the $\epsilon$-regime.  

By comparing the $\epsilon$-regime results with $p$-regime results
in a finite volume, we have also speculated on the regimes
of validity of the two approaches. It appears that for semi-realistic
lattices with a spatial extent of about 2~fm, the $\epsilon$-regime
approach might be applicable for $\mu = m \Sigma V \lsim 2.0$ 
and the $p$-regime for $ML \gsim 2.0$. In any case, 
the conventional infinite-volume formulae
are accurate (with errors below 10 -- 20\%) only at $ML \gsim 5.0$.

Finally, we have briefly addressed the effect of quenching on 
the determination of the $\Delta I = 1/2$ observables. 
New unphysical couplings are in general expected in the effective 
chiral theory with respect to the unquenched situation. 
In the $\epsilon$-regime we find, however, that the contamination from 
these new couplings is minimal at NLO: only one additional coupling 
enters our predictions, and we have shown that it is in principle possible 
to determine the quenched $g_8$ in spite of these quenching artifacts.

\section*{Acknowledgements}

This work is part of a bigger effort  whose goal
is to extract low-energy constants of QCD from numerical 
simulations with Ginsparg-Wilson fermions. 
We are indebted to our collaborators on this,  
L.~Giusti, C.~Pena, J.~Wennekers and H.~Wittig, for useful comments. 
The basic ideas of the general approach were developed 
in collaboration with M. L\"uscher and P.~Weisz; we would like to 
thank them for their input and for many valuable suggestions. 
P.~H.\ was supported in part by the Spanish CICYT 
(Project No.\ FPA2004-00996 and FPA2005-01678) and  
by the Generalitat Valenciana (Project No.\ GVA05/164). 


\appendix
\renewcommand{\thesection}{Appendix~\Alph{section}}
\renewcommand{\thesubsection}{\Alph{section}.\arabic{subsection}}
\renewcommand{\theequation}{\Alph{section}.\arabic{equation}}


\newpage

%
\section{Irreducible representations of the valence group}
\la{app:projections}

For completeness, we reiterate in this Appendix the main formulae
related to irreducible representations of the valence group SU($\Nv)$, 
relevant for the operators appearing in the weak Hamiltonian. 
We follow the tensor method discussed, e.g., in Ref.~\cite{hg2}. 

Like in the main body of the text, we make a distinction 
between the valence group SU($\Nv)$, used to classify the weak 
operators, and the full flavour symmetry SU($\Nf)$.
The indices $\ga,\gb,\gc,\gd,\ta,\tb,\tc,\td,\ha,\hb,\hc,\hd$ are assumed to 
take values in the valence subgroup only. We denote by
$O_{\ga\gb\gc\gd}$ a generic operator transforming under 
${\bf \Nv^*} \otimes {\bf \Nv^*} \otimes {\bf \Nv} \otimes {\bf \Nv}$
of SU($\Nv)$, and by $O_{\ta\tc}'$ one transforming under 
${\bf \Nv^*} \otimes {\bf \Nv}$.  

We define the projection operators 
\ba
 & & (P_1^\sigma)_{\ga\gb\gc\gd ; \ta\tb\tc\td} \equiv 
 \fr14 (\delta_{\ga\ta} \delta_{\gb\tb} + 
 \sigma \delta_{\ga\tb}\delta_{\gb\ta})
 (\delta_{\gc\tc} \delta_{\gd\td} + 
 \sigma \delta_{\gc\td} \delta_{\gd\tc})
 \;, \la{P1} \\
 & &  (P_2^\sigma)_{\ga\gb\gc\gd;\ta\tb\tc\td} \equiv  
 \delta_{\ga\ta} \delta_{\gb\tb} \delta_{\gc\tc} \delta_{\gd\td} 
 + \frac{1}{(\Nv + 2\sigma)(\Nv + \sigma)}
 ( 
 \delta_{\ga\gc} \delta_{\gb\gd} + \sigma
 \delta_{\ga\gd} \delta_{\gb\gc}) \delta_{\ta\tc} \delta_{\tb\td}
 \nn
 & & ~~~~~~~~ - \frac{1}{\Nv + 2\sigma}
 (
 \delta_{\ga\gc} \delta_{\gb\tb}  \delta_{\gd\td} \delta_{\ta\tc}+
 \delta_{\gb\gd} \delta_{\ga\ta} \delta_{\gc\tc} \delta_{\tb\td}+
 \sigma \delta_{\ga\gd} \delta_{\gb\tb} \delta_{\gc\td} \delta_{\ta\tc}+
 \sigma \delta_{\gb\gc} \delta_{\ga\ta} \delta_{\gd\tc} \delta_{\tb\td}
 )\;, \hspace*{1.0cm} \la{P2} \\
 & & (P_3)_{\ga\gc;\ta\tc} \equiv
 \delta_{\ga\ta}\delta_{\gc\tc} - 
 \frac{1}{\Nv} \delta_{\ga\gc}\delta_{\ta\tc} 
 \la{P3} \;.
\ea
In addition, $\Pv$ is defined to project from SU($\Nf$) to SU($\Nv$).
The operators denoted by $O_{27}$, $O_8$ and $O_{8}'$ 
can now be defined as
\ba
 {[O_{27}]}_{\ga\gb\gc\gd} & \equiv & 
 (P_2^+ P_1^+)_{\ga\gb\gc\gd;\ta\tb\tc\td}\, O_{\ta\tb\tc\td}^{\mbox{ }}
 \;, \la{o27proj} \\ 
 {[O_{8}^\pm]}_{\ga\gc} & \equiv &
 {(P_3)}_{\ga\gc;\ha\hc} 
 {(P_1^\pm)}_{\ha\hb\hc\hb ; \ta\tb\tc\td}\, O_{\ta\tb\tc\td}^{\mbox{ }}
 \;, \la{o8proj} \\
 {[O_{8}']}_{\ga\gc} & \equiv & 
 {(P_3)}_{\ga\gc;\ta\tc}\, O_{\ta\tc}'
 \;.
\ea
Note that the contraction over $\hb$ in \eq\nr{o8proj}
goes over valence flavours only, and that 
additional octet operators ($O_8^-$) can appear already 
at the leading order when $\Nv\neq\Nf$.

Instead of a generic operator $O_{\ga\gb\gc\gd}$, practical
computations of the type in Ref.~\cite{p4} 
involve certain factorised forms, like
\ba
 {[O_1]}_{\ga\gb\gc\gd} & \equiv & (Q)_{\gc\ga} (R)_{\gd\gb}
 \;, \\
 {[O_2]}_{\ga\gb\gc\gd} & \equiv & (Q)_{\gc\gb} (R)_{\gd\ga}
 \;, \\
 {[O_3]}_{\ga\gb\gc\gd} & \equiv & \delta_{\gc\ga} (R)_{\gd\gb}
 \;, \\
 {[O_4]}_{\ga\gb\gc\gd} & \equiv & \delta_{\gc\gb} (R)_{\gd\ga}
 \;, \\
 {[O_5]}_{\ga\gb\gc\gd} & \equiv & \delta_{\gc\ga} \delta_{\gd\gb}
 \;, \\
 {[O_6]}_{\ga\gb\gc\gd} & \equiv & \delta_{\gc\gb} \delta_{\gd\ga}
 \;.
\ea
Then projections of the types in \eqs\nr{o27proj}, \nr{o8proj} produce
\be
 \begin{array}{cclcccl}
 {[ P_2^\sigma P_1^\sigma O_1^{\mbox{ }}]}_{\ga\gb\gc\gd} & = & 
 S^\sigma_{\ga\gb\gc\gd}(Q,R)
 \;, & \hspace*{1cm} &
 {[ P^{\mbox{ }}_3 P_1^\sigma O^{\mbox{ }}_1]}_{\ga\gc} & = & 
 T^\sigma_{\ga\gc}(Q,R)
 \;, \\
 {[ P_2^\sigma P_1^\sigma O_2^{\mbox{ }}]}_{\ga\gb\gc\gd} & = & 
 \sigma S^\sigma_{\ga\gb\gc\gd}(Q,R)
 \;, & \hspace*{1cm} &
 {[ P^{\mbox{ }}_3 P_1^\sigma O^{\mbox{ }}_2]}_{\ga\gc} & = & 
 \sigma T^\sigma_{\ga\gc}(Q,R)
 \;, \\
 {[ P_2^\sigma P_1^\sigma O_3^{\mbox{ }}]}_{\ga\gb\gc\gd} & = & 
 0
 \;, & \hspace*{1cm} &
 {[ P^{\mbox{ }}_3 P_1^\sigma O^{\mbox{ }}_3]}_{\ga\gc} & = & 
 U^\sigma_{\ga\gc}(R)
 \;, \\
 {[ P_2^\sigma P_1^\sigma O_4^{\mbox{ }}]}_{\ga\gb\gc\gd} & = & 
 0
 \;, & \hspace*{1cm} &
 {[ P^{\mbox{ }}_3 P_1^\sigma O^{\mbox{ }}_4]}_{\ga\gc} & = & 
 \sigma U^\sigma_{\ga\gc}(R)
 \;, \\
 {[ P_2^\sigma P_1^\sigma O_5^{\mbox{ }}]}_{\ga\gb\gc\gd} & = & 
 0
 \;, & \hspace*{1cm} &
 {[ P^{\mbox{ }}_3 P_1^\sigma O^{\mbox{ }}_5]}_{\ga\gc} & = & 
 0
 \;, \\
 {[ P_2^\sigma P_1^\sigma O_6^{\mbox{ }}]}_{\ga\gb\gc\gd} & = & 
 0
 \;, & \hspace*{1cm} &
 {[ P^{\mbox{ }}_3 P_1^\sigma O^{\mbox{ }}_6]}_{\ga\gc} & = & 
 0
 \;,
 \end{array}
\ee
where
(introducing the notation $\trp(...) \equiv \tr(\Pv\, ...)$)
\ba
 S^\sigma_{\ga\gb\gc\gd}(Q,R) & = &
 \fr14
 \biggl\{
   Q_{\gc\ga}^{\mbox{ }} R_{\gd\gb}^{\mbox{ }} + 
   R_{\gc\ga}^{\mbox{ }} Q_{\gd\gb}^{\mbox{ }} + 
   \sigma ( 
   Q_{\gc\gb}^{\mbox{ }} R_{\gd\ga}^{\mbox{ }} + 
   R_{\gc\gb}^{\mbox{ }} Q_{\gd\ga}^{\mbox{ }}  
   )
 - \nn & & - 
 \frac{\delta^{\mbox{ }}_{\gd\ga}}{\Nv + 2 \sigma}
 \Bigl[
  (Q\Pv R + R \Pv Q)_{\gc\gb} + 
  \sigma 
  \Bigl(
     Q_{\gc\gb}^{\mbox{ }} \trp(R) + R_{\gc\gb}^{\mbox{ }} \trp(Q)
  \Bigr) \Bigr]
 - \nn & & - 
 \frac{\delta^{\mbox{ }}_{\gc\gb}}{\Nv + 2 \sigma}
 \Bigl[ 
  (Q\Pv R + R \Pv Q)_{\gd\ga} + 
  \sigma  
  \Bigl(
     Q_{\gd\ga}^{\mbox{ }} \trp(R) + R_{\gd\ga}^{\mbox{ }} \trp(Q)
  \Bigr)
  \Bigr] 
 - \nn & & - 
 \frac{\sigma\delta^{\mbox{ }}_{\gc\ga}}{\Nv + 2 \sigma}
 \Bigl[ 
  (Q\Pv R + R \Pv Q)_{\gd\gb} + 
  \sigma
  \Bigl(
     Q_{\gd\gb}^{\mbox{ }} \trp(R) + R_{\gd\gb}^{\mbox{ }} \trp(Q)
  \Bigr)
  \Bigr]
 - \nn & & - 
 \frac{\sigma\delta^{\mbox{ }}_{\gd\gb}}{\Nv + 2 \sigma}
  \Bigl[
  (Q\Pv R + R \Pv Q)_{\gc\ga}  +
  \sigma
  \Bigl(
     Q_{\gc\ga}^{\mbox{ }} \trp(R) + R_{\gc\ga}^{\mbox{ }} \trp(Q)
  \Bigr)
  \Bigr] 
 + \nn & & + 
 \frac{2(\delta^{\mbox{ }}_{\gc\gb} \delta^{\mbox{ }}_{\gd\ga} + 
         \delta^{\mbox{ }}_{\gc\ga} \delta^{\mbox{ }}_{\gd\gb})
   }{(\Nv + \sigma)(\Nv + 2 \sigma)}
 \Bigl[
  \trp(Q \Pv R) + \sigma \trp(Q) \trp(R) 
 \Bigr]
 \biggr\}
 \;, \la{Ssigma} \\ 
 T^\sigma_{\ga\gc}(Q,R) & = & 
 \fr14 
 \Bigl[
  \sigma(Q \Pv R + R \Pv Q)_{\gc\ga} 
 + Q_{\gc\ga}^{\mbox{ }} \trp(R) + R_{\gc\ga}^{\mbox{ }} \trp(Q) 
 \Bigr] - \nn
 & & - \frac{\delta_{\gc\ga}}{2\Nv}
 \Bigl[ 
  \trp(Q) \trp(R) + \sigma \trp(Q \Pv R)
 \Bigr] 
 \;, \\  
 U^\sigma_{\ga\gc}(R) & = & 
 \fr14 (\Nv + 2 \sigma) 
 \Bigl[  R_{\gc\ga}^{\mbox{ }}  -
 \frac{\delta_{\gc\ga}}{\Nv} 
 \trp(R)  
 \Bigr]
 \;. 
\ea
Considering, in particular, the operators
\ba
 \Delta^{(1)}_{\ga\gb\gc\gd} & = & 
 T^{\{a}_{\gc\ga} T^{b\}}_{\gd\gb}
 \;, \\ 
 \Delta^{(2)}_{\ga\gb\gc\gd} & = & 
 T^{\{a}_{\gc\gb} T^{b\}}_{\gd\ga} - \fr12
 \Bigl( 
   \delta^{\mbox{ }}_{\gc\gb} \{ T^a,T^b \}_{\gd\ga} + 
   \delta^{\mbox{ }}_{\gd\ga} \{ T^a,T^b \}_{\gc\gb}    
 \Bigr)
 \;, \\ 
 \Delta^{(3)}_{\ga\gb\gc\gd} & = & 
 T^{\{a}_{\gc\gb} T^{b\}}_{\gd\ga} + 
   \delta^{\mbox{ }}_{\gc\gb} \{ T^a,T^b \}_{\gd\ga} + 
   \delta^{\mbox{ }}_{\gd\ga} \{ T^a,T^b \}_{\gc\gb}    
 \;, \\ 
 \Delta^{(4)}_{\ga\gb\gc\gd} & = & 
   \delta^{\mbox{ }}_{\gc\ga} \{ T^a,T^b \}_{\gd\gb} + 
   \delta^{\mbox{ }}_{\gd\gb} \{ T^a,T^b \}_{\gc\ga}    
  \;,
\ea
which appear in the computations of \fig\ref{fig:graphs}, and choosing
the indices that appear in the physical 
operators $O_{27}$ and $O_{8}$, we obtain
\be
 \begin{array}{cclcccl}
 {[ P_2^+ P_1^+ \Delta^{(1)}]}_{suud} & = & 
 2 S^+_{suud}(T^a,T^b)
 \;, & \hspace*{0.5cm} &
 {[ P^{\mbox{ }}_3 P_1^{\sigma} \Delta^{(1)}]}_{sd} & = & 
 \fr12\{T^a,T^b\}_{ds} \, \sigma
 \;, \\
 {[ P_2^+ P_1^+ \Delta^{(2)}]}_{suud} & = & 
 2 S^+_{suud}(T^a,T^b) 
 \;, & \hspace*{0.5cm} &
 {[ P^{\mbox{ }}_3 P_1^{\sigma} \Delta^{(2)}]}_{sd} & = & 
 \fr12\{T^a,T^b\}_{ds} \Bigl( - \frac{\sigma}{2} \Nv \Bigr) 
 \;, \\
 {[ P_2^+ P_1^+ \Delta^{(3)}]}_{suud} & = & 
 2 S^+_{suud}(T^a,T^b)
 \;, & \hspace*{0.5cm} &
 {[ P^{\mbox{ }}_3 P_1^{\sigma} \Delta^{(3)}]}_{sd} & = & 
 \fr12\{T^a,T^b\}_{ds} (\sigma \Nv + 3 )
 \;, \\
 {[ P_2^+ P_1^+ \Delta^{(4)}]}_{suud} & = & 
 0
 \;, & \hspace*{0.5cm} &
 {[ P^{\mbox{ }}_3 P_1^{\sigma} \Delta^{(4)}]}_{sd} & = & 
 \fr12\{T^a,T^b\}_{ds} ( \Nv + 2 \sigma )
 \;.
 \end{array} \la{Tproj}
\ee
For $\Nv = 3$, 
the function $\Delta^{ab}_{27}\equiv 2 S^+_{suud}(T^a,T^b)$ 
is shown explicitly in \eq\nr{flavproj}.

%
\section{Ultraviolet divergences and $\rmO(p^4)$ operators}

Once we go beyond the order $\rmO(p^2)$ in \xpt, 
the number of operators that enter \eq\nr{Lw_XPT} increases
dramatically. At the order $\rmO(p^4)$, we rewrite the  
weak Hamiltonian as
\be
  {\cal H}_w \equiv  2 \sqrt{2} G_F V_{ud} V^*_{us}
  \biggl\{ 
  \fr53 \Bigl[ g_{27} {\cal O}_{27} 
  + \sum_i D_i \bar\mathcal{O}_{27}^{(i)} \Bigr]
  + 2 \Bigl[ g_8 {\cal O}_8
  + g_8'{\cal O}'_8
  + \sum_i E_i \bar\mathcal{O}_8^{(i)} \Bigr]
  \biggr\} + \Hc  \,,
 \la{Lw_XPT_full}
\ee
where $\bar\mathcal{O}_{27}^{(i)}$, $\bar\mathcal{O}_{8}^{(i)}$
are the new operators. 
For $\Nf = \Nv = 3$, (over)complete sets for 
$\bar\mathcal{O}_{27}^{(i)}$, $\bar\mathcal{O}_{8}^{(i)}$ have
been listed in Ref.~\cite{p4}. 
The use of partial integration 
identities makes it possible to reduce the number of operators drastically, 
leading to the lists commonly used in phenomenology~\cite{p4reduced};
in our case, however, the use of partial integration identities 
is not possible, since we consider local operator insertions
(i.e. $\mathcal{H}_w$ is not integrated over spacetime). 

Generalizing \eq\nr{C1xpt}, we define the correlation functions
now with the LECs added, 
\ba
 {[\mathcal{C}_{27}]}^{ab} (x_0,y_0) & \equiv &   
 \int\! {\rm d}^3x
 \int\! {\rm d}^3y\, \Bigl\langle \mathcal{J}^a_0(x) 
 \Bigl[ g_{27} {\cal O}_{27}(0)  
  + \sum_i D_i \bar\mathcal{O}_{27}^{(i)}(0) \Bigr]
 \mathcal{J}^b_0(y) 
 \Bigr \rangle 
 \;, \\
 {[\mathcal{C}_{8}]}^{ab} (x_0,y_0) & \equiv &   
 \int\! {\rm d}^3x
 \int\! {\rm d}^3y\, \Bigl\langle \mathcal{J}^a_0(x) 
 \Bigl[ 
    g_8 {\cal O}_8(0)
  + g_8'{\cal O}'_8(0)
  + \sum_i E_i \bar\mathcal{O}_8^{(i)} (0)
 \Bigr]
 \mathcal{J}^b_0(y) 
 \Bigr \rangle 
 \;. \la{C1xpt_full} \hspace*{1.0cm}
\ea
The results can be written in the forms 
\ba
  {[
  \mathcal{C}_{27}
  ]}^{ab}(x_0,y_0) 
 \!\! & = & \!\!  
 \Delta^{ab}_{27}
 \biggl\{ g_{27} \Bigl[  \mathcal{C}(x_0) \mathcal{C}(y_0) 
 +  \mathcal{D}_{27}(x_0,y_0)  \Bigr] 
 +  \mathcal{E}_{27} (x_0,y_0)
 \biggr\} 
 \;, \la{C27}  \\
  {[
  \mathcal{C}_{8}
  ]}^{ab}(x_0,y_0)
 \!\! & = & \!\!  
  \Delta^{ab}_8 \biggl\{ g_8 
 \Bigl[ \mathcal{C}(x_0) \mathcal{C}(y_0) 
 + \mathcal{D}_{8}(x_0,y_0)  \Bigr]  + 
 g_8' \mathcal{D}'_{8}(x_0,y_0)
 + \mathcal{E}_{8} (x_0,y_0)
 \biggr\}
 \;, \la{C8} \hspace*{0.5cm}
\ea
where $ \Delta^{ab}_{27}=
2 S^+_{suud}(T^a,T^b)$ in the notation
of \eq\nr{Tproj}, and $\Delta^{ab}_{8} = \{T^a,T^b \}_{ds}/2$.

%
The 
list of operators from Ref.~\cite{p4} (modulo certain minus-signs) 
that can contribute to $\mathcal{C}_{27}$ at NLO  is  constituted
by the properly projected (cf.\ Appendix A) versions of:
\ba
 {\bar{\mathcal O}}_{27}^{(2)} 
  &=& - (\mP)_{\gc\ga} (\mP)_{\gd\gb} 
 \;, \la{o27_2} \\ 
 {\bar {\mathcal O}}_{27}^{(4)} 
 &=& \left(\mL_\mu\right)_{\gc\ga}
 \left\{\mL_\mu,  \mS\right\}_{\gd\gb}
 \;, \\ 
 {\bar{\mathcal O}}_{27}^{(7)} 
 &=& \left(\mL_\mu\right)_{\gc\ga}
 \left(\mL_\mu \right)_{\gd\gb} \tr\left(\mS\right)
 \;, \\ 
 {\bar{\mathcal O}}_{27}^{(19)}
  &=& i\left(\mW_{\mu\mu}\right)_{\gc\ga} (\mP)_{\gd\gb}
 \;, \\ 
 {\bar{\mathcal O}}_{27}^{(20)} 
 &=& - \left(\mL_\mu\right)_{\gc\ga}
 \left(\partial_\nu \mW_{\mu\nu} \right)_{\gd\gb} 
 \;, \\ 
 {\bar{\mathcal O}}_{27}^{(21)}
  &=& - \left(\mL_\mu\right)_{\gc\ga}
 \left(\partial_\mu \mW_{\nu\nu} \right)_{\gd\gb} 
 \;, \\ 
 {\bar{\mathcal O}}_{27}^{(24)} 
 &=& - \left(\mW_{\mu\nu}\right)_{\gc\ga}
 \left(\mW_{\mu\nu} \right)_{\gd\gb} 
 \;, \\ 
 {\bar{\mathcal O}}_{27}^{(25)}
  &=& - \left(\mW_{\mu\mu}\right)_{\gc\ga}
 \left(\mW_{\nu\nu} \right)_{\gd\gb}\;. 
 \la{o27_25} 
\ea
Here we utilize the notation 
\ba
 \mS\!\!&\equiv   U \chi^\dagger +  \chi U^\dagger \;, \;\;\;\;\;
 \mP\!\!&\equiv  i\left( U \chi^\dagger -  \chi U^\dagger\right)
\;, \la{ops1} \\ 
 \mL_\mu \!\!\!\! &\equiv U  \partial_\mu U^\dagger 
\;, \;\;\;\;\;\;\;\;\; 
 \mW_{\mu\nu}\!\!\!\! &\equiv 2
 \left( \partial_\mu \mL_\nu + \partial_\nu \mL_\mu \right) 
 \;, \la{ops2} 
\ea
where $\chi\equiv {2 m \Sigma }/{ F^2} = M^2$. 
As stressed in Ref.~\cite{p4}, not all of these operators are 
independent, however: equations of motion can be used to eliminate 
19, 21, and 25, for instance.
In the following, we keep for generality 
all the operators.

The contribution from the $\rmO(p^4)$-constants to \eq\nr{C27} reads
\ba
 {\mathcal E}_{27}(x_0,y_0) &=& 4 M^4  P'(x_0) P'(y_0) 
 \left[D_2  +2 D_{19} 
 - 4 D_{24} - 4 D_{25} \right]  + \nonumber\\
 &+& 4 M^6 P(x_0) P(y_0)
 \left[D_4  + {\Nf\over 2} D_{7} - D_{20} - 
 {D_{21}} - F^2 (\Nf L_4 +L_5) \right] 
 \;. \hspace*{1cm}
 \label{eq:p4}
\ea
Taking into account that $\mathcal{C}(x_0)$ is finite; 
that $\mathcal{D}_{27}(x_0,y_0)$ contains the divergences
specified in \eq\nr{divD27}; 
that the QCD $\rmO(p^4)$ constants contain the divergence
($\lambda \equiv -1/32\pi^2\epsilon$)
\be
 \Nf L_4 + L_5 = \Nf L_4^r + L_5^r + \frac{\Nf}{4} \lambda 
 \;, \la{divL4}
\ee
where $L_4^r$, $L_5^r$ are finite; 
and that the $\rmO(p^4)$ constants contain the divergences
\ba
 D_4 &=& {D_4^r} + g_{27} \, F^2 \lambda  
 \Bigl( {\Nf+ 3 \over 8} \Bigr)
 \;, \nonumber\\ 
 D_7 &=& {D_7^r} + g_{27} \, F^2 \lambda 
 \Bigl( {1 \over 4} \Bigr)
 \;, \nonumber\\ 
 D_{20} &=& {D_{20}^r}  + g_{27} \, F^2 \lambda
 \Bigl( {1 \over 8} \Bigr)
 \;, \nonumber\\ 
 D_{24} &=& {D_{24}^r}  + g_{27} \, F^2 \lambda 
 \Bigl({1 \over 32} \Bigr) 
\;, \la{eq:duv}
\ea
the correlation function $\mathcal{C}_{27}$ in \eq\nr{C27}
can be seen to be finite.

%
As far as the octet correlation functions are concerned, 
it is the following types 
among the operators listed in Ref.~\cite{p4}
that contribute to the correlation function
$\mathcal{C}_8$ at the order we are considering:   
\ba
 \bar{\mathcal{O}}_{8}^{(1)} & \equiv & -( \mS \mS )_{ds} 
 \;, \la{bO81} \\ 
 \bar{\mathcal{O}}_{8}^{(2)} & \equiv & -(\mS)_{ds} \tr(\mS)
 \;, \la{bO82} \\ 
 \bar{\mathcal{O}}_{8}^{(3)} & \equiv & -(\mP \mP)_{ds}
 \;, \\ 
 \bar{\mathcal{O}}_{8}^{(10)} & \equiv & \{\mS,\mL_\mu \mL_\mu\}_{ds}
 \;, \\ 
 \bar{\mathcal{O}}_{8}^{(11)} & \equiv & (\mL_\mu \mS \mL_\mu)_{ds}
 \;, \\ 
 \bar{\mathcal{O}}_{8}^{(14)} & \equiv & (\mL_\mu \mL_\mu)_{ds} \tr(\mS)
 \;, \\ 
 \bar{\mathcal{O}}_{8}^{(33)} & \equiv & i\{\mW_{\mu\mu},\mP\}_{ds}
 \;, \\ 
 \bar{\mathcal{O}}_{8}^{(35)} & \equiv & -\{\mL_\mu,
                                            \partial_\nu\mW_{\mu\nu}\}_{ds}
 \;, \\ 
 \bar{\mathcal{O}}_{8}^{(36)} & \equiv & -\{\mL_\mu,
                                            \partial_\mu\mW_{\nu\nu}\}_{ds}
 \;, \\ 
 \bar{\mathcal{O}}_{8}^{(39)} & \equiv & -(\mW_{\mu\nu} \mW_{\mu\nu})_{ds}
 \;, \\ 
 \bar{\mathcal{O}}_{8}^{(40)} & \equiv & -(\mW_{\mu\mu} \mW_{\nu\nu})_{ds}
 \;. \la{bO840}
\ea
Again, there are relations between these operators: 
equations of motion can be used to eliminate 33, 36 and 40~\cite{p4}.
For $\Nf=\Nv$ the contributions from the QCD and weak 
$\rmO(p^4)$-constants to \eq\nr{C8} read
\ba
 {\mathcal E}_{8}(x_0,y_0) 
 \!\!\! &=& \!\!\! 8 M^4 P'(x_0) P'(y_0)
\Bigl[
 -E_1 -\frac{\Nf}{2} E_2 + E_3 
 + 4 E_{33} - 4 E_{39} - 4 E_{40}
 \Bigr]\! +  \nonumber\\
 &+& \!\!\! 8 M^6 P(x_0) P(y_0)
\Bigl[ 
 E_{10} + \fr12 E_{11} 
 + \frac{\Nf}{2} E_{14}  
  - 2 E_{35} - 2 E_{36}
 \Bigr] + 
 \nonumber\\[2mm]
 &+& \!\!\! 4 M^6 P(x_0) P(y_0)
 g_8 F^2 \left[ 
  -  \Nf L_4 - L_5 \right] + 
 \nonumber\\[2mm] 
 &+& \!\!\!
 4  M^6 \frac{{\rm d}}{{\rm d} M^2} \Bigl[ P'(x_0) P'(y_0) \Bigr] 
 g_8' F^2 \Bigl[ -\Nf L_4 - L_5 + 2 (\Nf L_6 + L_8 )\Bigr]
 \;.  \la{eq:ep4}
\ea
The results for the divergent parts of $E_i$ 
can be found in Ref.~\cite{p4} for $\Nf=3$ 
and will be given below for general $\Nf$.
Taking into account that (in the unquenched case)
\ba
 \Nf L_4 + L_5 - 2 (\Nf L_6 + L_8 ) & = & 
 \Nf L_4^r + L_5^r - 2 (\Nf L_6^r + L_8^r ) + \frac{\lambda}{4\Nf}
 \;, \la{divL6}
\ea
where $L_6^r$, $L_8^r$ are finite, and summing together with 
the divergences shown in \eqs\nr{divD8} and \nr{divD8p}, 
it can be verified that $\mathcal{C}_8$ is finite. 

%
\section{The case $\Nf\neq \Nv$}
 
For $\Nf \neq \Nv$, the set of possible operators is in general 
larger than for $\Nf = \Nv$: 
the only restrictions are that the operators be singlets
in the full group SU($\Nf)_R$, and have the correct transformation 
properties in the subgroup SU($\Nv)_L$. 
At $\rmO(p^2)$ this does not change the situation for the 27-plet, 
but it increases the amount of octets to four in total. 
Besides $\mathcal{O}_{8}$ defined by \eqs\nr{formofR}, 
\nr{O_XPT}, \nr{preO8}, 
{\it viz.}
\ba
 \mathcal{O}_8 & = & \frac{F^4}{8}
 \Bigl[ 
   \left(\mL_\mu \Pv \mL_\mu\right)_{ds} 
 + \left(\mL_\mu\right)_{ds}  \tr (\Pv \mL_\mu)
 \Bigr] 
 \;, 
\ea
and $\mathcal{O}'_{8}$ defined by \eq\nr{formofO8p}, 
there are two additional octets, which we choose to define 
such that they vanish in the limit $\Nf\rightarrow \Nv$:
\ba
 \hat{\mathcal{O}}_8 & \equiv & \frac{F^4}{8}
 \bigl[  
   \mL_\mu  \left( 1 - \Pv \right) \mL_\mu\bigr]_{ds} 
 \;,  \\
 \check{\mathcal{O}}_8 & \equiv & \frac{F^4}{8}
 \left(\mL_\mu\right)_{ds} 
 \tr \bigl[  
   \left( 1 - \Pv \right) \mL_\mu\bigr]  
 \;.
\ea
It should also be noted that these operators only 
contribute starting at the NLO, since at tree-level
they do not couple to two valence-flavoured mesons.
Since for $\mathcal{C}_{27}$ nothing changes with respect to Appendix~B, 
we concentrate on the octets in the following. 

The three-point octet correlation function is now of the form
\ba
 {[\mathcal{C}_{8}]}^{ab}  \!\! & \equiv & \!\!   
 \int\! {\rm d}^3x \!
 \int\! {\rm d}^3y\, \Bigl\langle \mathcal{J}^a_0(x) 
 \Bigl[ 
    g_8 {\cal O}_8
  + g_8'{\cal O}'_8
  + \hat{g}_8 \hat{\cal O}_8
  + \check{g}_8\check{\cal O}_8
  + \sum_i E_i \bar\mathcal{O}_8^{(i)} 
 \Bigr](0)
 \mathcal{J}^b_0(y) 
 \Bigr \rangle 
 \hspace*{1.0cm} \\ 
 & = & \!\!
  \Delta^{ab}_8 \biggl\{ g_8 
 \Bigl[ \mathcal{C}(x_0) \mathcal{C}(y_0) 
 + \mathcal{D}_{8}(x_0,y_0)  \Bigr]  
 + g_8' \mathcal{D}'_{8}
 + \hat{g}_8 \hat{\mathcal{D}}_{8}
 + \check{g}_8 \check{\mathcal{D}}_{8}
 + \mathcal{E}_{8} 
 \biggr\}
 \;, \la{C8nv} \hspace*{0.5cm}
\ea
where $\mathcal{D}_8$ can be found in \eq\nr{D8} 
and $\mathcal{D}_8'$ in \eq\nr{calDp}. The new functions read
\ba
 \hat{\mathcal{D}}_8(x_0,y_0) & = & 
 (\Nf - \Nv)
 \Bigl[
  -\fr12 \mathcal{D}_{27}(x_0,y_0) + \frac{F^2 M^2}{8} 
  \mathcal{I}_A(x_0,y_0) 
 \Bigr]
 \;, \la{D8pp} \\
 \check{\mathcal{D}}_8(x_0,y_0) & = & 
 \frac{F^2 M^2}{4}
 \Bigl[
  \Nv \mathcal{I}_B (x_0,y_0) - \mathcal{I}_A(x_0,y_0) 
 \Bigr]
 \;, \la{D8ppp} 
\ea
where we have defined 
\ba
 \mathcal{I}_A(x_0,y_0) & \equiv & 
 G(0;M^2) P'(x_0) P'(y_0) 
 - \nn & & \hspace*{0.0cm} -
 \frac{M^2}{2} P(x_0 - y_0)
  \Bigl[
   B(x_0) + B(y_0) 
  \Bigr] 
 + \nn & & \hspace*{0.0cm} 
  + \fr12 P'(x_0 - y_0)
   \Bigl[ B'(x_0) - B'(y_0) 
   \Bigr] 
 + \nn & & \hspace*{0.0cm} 
 + M^4 \int_0^T \! {\rm d}  \tau \,
 B(\tau) P(\tau - x_0) P(\tau - y_0)  
  \;, \la{IA} \\  
 \mathcal{I}_B(x_0,y_0) & \equiv &
 E(0;M^2) P'(x_0) P'(y_0) 
 - \nn & & \hspace*{0.0cm} -
   \frac{M^2}{2} P(x_0 - y_0)
  \Bigl[
   \tilde B(x_0) + \tilde B(y_0) 
  \Bigr] 
 + \nn & & \hspace*{0.0cm} 
   + \fr12 P'(x_0 - y_0)
   \Bigl[
    \tilde B'(x_0) + \tilde B_0(x_0)  
    - \tilde B'(y_0) - \tilde B_0(y_0) \Bigr] 
 + \nn & & \hspace*{0.0cm} +
  M^2 \int_0^T \! {\rm d}  \tau \,
 \Bigl[ M^2 \tilde B(\tau) +\fr12 \tilde B_{00}(\tau) \Bigr] 
 P(\tau - x_0) P(\tau - y_0)
 \;, \la{IB}
\ea
and the notation follows that in \eq\nr{D8}.
The divergent parts read (in the unquenched case) 
\ba
 \hat{\mathcal{D}}_{8}(x_0,y_0) & = &  \hat{\mathcal{D}}_{8}^r(x_0,y_0)
 + \frac{F^2 \lambda}{4} (\Nf - \Nv)
 \Bigl[ M^6 P(x_0) P(y_0)
 \Bigr]
 \;, \la{divD8pp} \\
 \check{\mathcal{D}}_{8}(x_0,y_0) & = &  \check{\mathcal{D}}_{8}^r(x_0,y_0)
 + \frac{F^2 \lambda}{2} \Bigl( 1 - \frac{\Nv}{\Nf} \Bigr)
 \Bigl[  M^6 P(x_0) P(y_0)  -  
  M^4 P'(x_0) P'(y_0)
 \Bigr]
 \;, \hspace*{0.9cm} \la{divD8ppp} 
\ea
where $\hat{\mathcal{D}}_{8}^r$, $\check{\mathcal{D}}_{8}^r$ are 
finite. 

The list of operators contributing to $\mathcal{E}_8$ 
for $\Nf\neq\Nv$ is also much longer. 
We will not provide any systematic classification of all the 
possibilities, but only list the additional operators that are needed for 
cancelling the ultraviolet divergences at NLO. Using the same notation
as in Appendix A [$\trp(...) \equiv \tr(\Pv\, ...)$], we need 
\ba
 \bar{\mathcal{O}}_{8}^{(1')} & \equiv & -( \mS \Pv \mS )_{ds} 
 \;, \\ 
 \bar{\mathcal{O}}_{8}^{(2')} & \equiv & -(\mS)_{ds} \trp(\mS)
 \;, \\ 
 \bar{\mathcal{O}}_{8}^{(10')} & \equiv & \{\mS,\mL_\mu \Pv \mL_\mu\}_{ds}
 \;, \\ 
 \bar{\mathcal{O}}_{8}^{(10'')} & \equiv & 
    (\mS \Pv \mL_\mu \mL_\mu + \mL_\mu \mL_\mu \Pv \mS )_{ds}
 \;, \\ 
 \bar{\mathcal{O}}_{8}^{(11')} & \equiv & 
 \fr12 (\mL_\mu \{\Pv, \mS\} \mL_\mu)_{ds}
 \;, \\ 
 \bar{\mathcal{O}}_{8}^{(14')} & \equiv & (\mL_\mu \Pv \mL_\mu)_{ds} \tr(\mS)
 \;, \\ 
 \bar{\mathcal{O}}_{8}^{(14'')} & \equiv & (\mL_\mu \mL_\mu)_{ds} \trp(\mS)
 \;, \\ 
 \bar{\mathcal{O}}_{8}^{(35')} & \equiv & 
  -(\mL_\mu \Pv \partial_\nu\mW_{\mu\nu} + 
     \partial_\nu\mW_{\mu\nu} \Pv \mL_\mu )_{ds}
 \;, \\ 
 \bar{\mathcal{O}}_{8}^{(39')} & \equiv & -(\mW_{\mu\nu} \Pv \mW_{\mu\nu})_{ds}
 \;. 
\ea
With these definitions, we get:
\ba
 {\mathcal E}_{8}(x_0,y_0) 
 \!\!\! &=& \!\!\! 8 M^4 P'(x_0) P'(y_0)
\Bigl[
 -E^{}_1 -E^{}_{1'} -\frac{\Nf}{2} E^{}_2 -\frac{\Nv}{2} E^{}_{2'} + E^{}_3 
 + \nn
 && \hphantom{ 4 M^4 P'(x_0) P'(y_0) }
 + 4 E^{}_{33} - 4 (E^{}_{39} + E^{}_{39'}) - 4 E^{}_{40}
 \Bigr]\! +  \nonumber\\
 &+& \!\!\! 8 M^6 P(x_0) P(y_0)
\Bigl[ 
 E^{}_{10} + E^{}_{10'} + E^{}_{10''} +  
 \fr12 (E^{}_{11} + E^{}_{11'}) 
 + 
 \nn 
 && \hphantom{ 4 M^6 P(x_0) P(y_0) }
 + \frac{\Nf}{2} (E^{}_{14} + E^{}_{14'} ) + 
 \frac{\Nv}{2} E^{}_{14''}
  - 2 (E^{}_{35} + E^{}_{35'} ) - 2 E^{}_{36}
 \Bigr] + 
 \nonumber\\[2mm]
 &+&  \!\!\! 4 M^6 P(x_0) P(y_0)
 g^{}_8 F^2 \left[ 
  -  \Nf L_4 - L_5 \right]  + 
  \nonumber\\[2mm] 
  &+& \!\!\!
  4  M^6 \frac{{\rm d}}{{\rm d} M^2} \Bigl[ P'(x_0) P'(y_0) \Bigr] 
  g_8' F^2 \Bigl[ -\Nf L_4 - L_5 + 2 (\Nf L_6 + L_8 )\Bigr]
  \;.  \la{eq:ep4nv}
\ea

Like for $\mathcal{C}_{27}$, 
the part $\mathcal{C}(x_0)\mathcal{C}(y_0)$ 
in \eq\nr{C8nv} is finite, while the other parts
contain divergences. More precisely,  
$\mathcal{D}_8$, $\mathcal{D}_8'$, 
$\hat{\mathcal{D}}_8$, $\check{\mathcal{D}}_8$ are of the forms
shown in \eqs\nr{divD8}, \nr{divD8p}, \nr{divD8pp}, \nr{divD8ppp},  
the combination
$\Nf L_4 + L_5$ of the form in \eq\nr{divL4}, 
while the combination on the last line of \eq\nr{eq:ep4nv}
is of the form in \eq\nr{divL6}.
Moreover, writing
\be
 E_i = E_i^r 
 + \frac{F^2 \lambda}{2} 
 \Bigl( g_8 \eta_i 
  + g_8' \eta'_i
  + \hat{g}_8 \hat{\eta}_i
  + \check{g}_8 \check{\eta}_i \Bigr)
 \la{Edivs}
\ee
where $E_i^r$ are finite, 
the coefficients $\eta_i$, $\eta'_i$, $\hat{\eta}_i$, $\check{\eta}_i$ 
can be derived with the 
method of Ref.~\cite{p4}; they are listed in Table 1. 
Summing together, all the divergences cancel
in $\mathcal{C}_8$, as they should.

\begin{table}[ht]

\begin{center}
\begin{tabular}{lllll}
\hline\hline
 $i$ & $\eta_i$ & $\eta'_i$ & $\hat{\eta}_i$ & $\check{\eta}_i$ \\
\hline\hline
 ${1}$ &  
 $(\Nv + 2) \Bigl( \frac{1}{16} - \frac{1}{4\Nf} \Bigr)$ & 
 $ -\frac{\Nf}{4} + \frac{1}{\Nf}$ & 
 $ \frac{\Nf - \Nv}{16} - \frac{1}{4\Nf}$ & 
 $-\fr18 + \frac{\Nv}{4\Nf}$ \\
 ${1'}$ & 
 $\frac{1}{8} - \frac{1}{4\Nf}$ & 
 $0$ & 
 $\frac{1}{4\Nf}$ & 
 $-\fr18$ \\
 ${2}$ &
 $(\Nv + 2) \frac{1}{4\Nf^2}$ & 
 $-\frac{1}{4} - \frac{1}{2 \Nf^2}$ & 
 $\fr18$ & 
 $-\frac{\Nv}{4\Nf^2}$ \\
 ${2'}$ & 
 $\frac{1}{8} - \frac{1}{4\Nf}$ & 
 $0$ & 
 $-\fr18$ & 
 $\frac{1}{4\Nf}$ \\
 ${3}$ & 
 $0$ & 
 $0$ & 
 $0$ & 
 $0$ \\
 ${10}$ &
 $(\Nv + 2) \Bigl( - \frac{1}{32} + \frac{1}{16\Nf} \Bigr)$ & 
 $\frac{\Nf}{8}$ & 
 $\frac{\Nf + \Nv}{32}$ & 
 $\frac{1}{16}\Bigl( 1 - \frac{\Nv}{\Nf} \Bigr)$ \\
 ${10'}$ & 
 $\frac{3}{16} + \frac{\Nf}{16}$ & 
 $0$ & 
 $-\frac{\Nf}{16}$ & 
 $-\frac{3}{16}$ \\
 ${10''}$ & 
 $-\frac{3}{16}$ & 
 $0$ & 
 $0$ & 
 $\frac{3}{16}$ \\
 ${11}$ &
 $(\Nv + 2) \Bigl( - \frac{3}{8\Nf}  \Bigr)$ & 
 $0$ & 
 $\frac{\Nf}{8}$ & 
 $\frac{3\Nv}{8\Nf}$ \\
 ${11'}$ & 
 $\frac{3}{8} + \frac{\Nf}{8}$ &
 $0$ & 
 $-\frac{\Nf}{8}$ & 
 $-\fr38$ \\
 ${14}$ & 
 $0$ & 
 $0$ & 
 $\frac{1}{16}$ & 
 $0$ \\
 ${14'}$ &
 $\frac{1}{4}$ & 
 $0$ & 
 $-\fr14$ & 
 $0$ \\
 ${14''}$ &
 $-\frac{3}{16}$ &
 $0$ & 
 $\frac{3}{16}$ &
 $0$ \\
 ${33}$ & 
 $(\Nv + 2) \Bigl( \frac{1}{64} - \frac{1}{32\Nf}  \Bigr)$ &  
 $0$ & 
 $\frac{\Nf - \Nv}{64}$ & 
 $-\frac{1}{32}\Bigl( 1 - \frac{\Nv}{\Nf}\Bigr)$ \\
 ${35}$ & 
 $(\Nv + 2) \Bigl( - \frac{1}{32}\Bigr)$ & 
 $0$ & 
 $\frac{\Nv - \Nf}{32}$ & 
 $\frac{1}{16}$ \\
 ${35'}$ & 
 $\frac{1}{16}$ & 
 $0$ & 
 $0$ & 
 $-\frac{1}{16}$ \\
 ${36}$ & 
 $0$ & 
 $0$ & 
 $0$ & 
 $0$ \\
 ${39}$ & 
 $(\Nv + 2) \Bigl( - \frac{1}{64} \Bigr)$ & 
 $0$ & 
 $\frac{\Nv - \Nf}{64}$ & 
 $\frac{1}{32}$ \\
 ${39'}$ &
 $\frac{1}{32}$ & 
 $0$ & 
 $0$ & 
 $-\frac{1}{32}$ \\
 ${40}$ &
 $0$ & 
 $0$ & 
 $0$ & 
 $0$ \\
\hline\hline
\end{tabular}
\end{center}

\caption[a]{\small The coefficients that appear in \eq\nr{Edivs}, 
in the unquenched case.}

\end{table}

The results for $E_i$ that need to be used for the case $\Nf =\Nv$ 
can be obtained from Table~1 by summing together the 
coefficients with the same ``numerical'' index that 
then correspond to the coefficients of 
the operators in \eqs\nr{bO81}--\nr{bO840}:
$(E_{1}+E_{1'})_{\Nf = \Nv}$ for \eq\nr{bO81}, 
$(E_{2}+E_{2'})_{\Nf = \Nv}$ for \eq\nr{bO82}, etc.
It can immediately be seen that the divergent parts proportional to 
$\hat{g}_8$ and $\check{g}_8$ cancel in these sums, as has to be the case. 

We finally comment on the quenched limit, 
corresponding formally to $\Nf\to 0$ but $\Nv$ fixed.
We have seen that for $\Nf\neq\Nv$ additional operators 
in general appear, as elaborated in Ref.~\cite{gp}. 
However, it is easy to see that 
the functions $\mathcal{I}_A$, $\mathcal{I}_B$
that appear in \eqs\nr{D8pp}, \nr{D8ppp}, 
vanish in the $\epsilon$-regime. Therefore the coefficient
$\check{g}_8$ does not contribute in \eq\nr{C8nv} in the $\epsilon$-regime. 
Moreover, $\hat{\mathcal{D}}_8$ is determined by the same 
function $\mathcal{D}_{27}$ that appears in 
$\mathcal{D}_8$ (cf.\ \eqs\nr{D8}, \nr{D8pp}). 
In particular, the normalised 
three-point function defined in analogy
with \eq\nr{ratio8} obtains for $\Nf\to 0$ the form
\be
 \Delta^{ab}_8\biggl[ g_8 - (g_8 - \hat{g}_8) 
 \frac{\Nv}{(FL)^2}
 \Bigl( \rho^{-\fr12} {\beta_1}  - \rho\, k_{00} 
 \Bigr)
 \biggr]
 + g_8' {[\mathcal{C}_{8}']}^{ab}_\rmi{norm}(x_0,y_0)
 \;.  \la{O8quench}
\ee
We observe that quenched functional behaviour 
only appears in the part ${[\mathcal{C}_{8}']}^{ab}_\rmi{norm}(x_0,y_0)$ 
(cf.\ \eq\nr{qC8p}), and can thus
be eliminated by disentangling the contributions related to $g_8'$, 
just like in Sec.~\ref{ss:o8eps}. Moreover, it can be verified
that at the NLO in the $\epsilon$-regime, 
the coefficients $\hat{g}_8$, $\check{g}_8$ do not contribute
to the correlation function considered in Sec.~\ref{ss:direct},
such that $g_8'$ can be separately determined just like there. 
The remaining terms in~\eq\nr{O8quench}
can be disentangled in principle by monitoring the volume 
dependence, from which it should be possible to determine
the ``physical'' coefficient $g_8$.

%
\section{Correlation functions for $\Nv = 4$}

It has been argued recently that many of the mysteries related
to the $\Delta I = 1/2$ rule can be studied particularly cleanly
[both from the conceptual and from the practical point of view] 
by considering the SU(4) symmetric situation, 
i.e. $\Nv = \Nf = 4$~\cite{strategy}.
We discuss here how our predictions can be converted to apply
to that situation. 

Rather than 27, 8, the dimensions of the relevant irreducible 
representations are 84, 20 for $\Nv = 4$. The corresponding 
operators are obtained like the 27 for $\Nv = 3$, but by using
the projection operators $P_2^\sigma P_1^\sigma$ in \eq\nr{o27proj},
with $\sigma = +1$ for the 84 and $\sigma = -1$ for the 20.
Following the notation in Ref.~\cite{strategy}, 
the corresponding operators are denoted by 
$[\hat \mathcal{O}_{1}]_{\a\b\c\d}^\sigma$.

The three-point correlation function we are interested in now 
takes the form 
\be
 {[\hat \mathcal{C}_{1}]}^{ab,\sigma}_{\a\b\c\d}(x_0,y_0) \equiv  
 \int\! {\rm d}^3x\, 
 \int\! {\rm d}^3y\, \Bigl\langle \mathcal{J}^a_0(x) 
 \Bigl\{
    g_1^{\sigma} [\hat\mathcal{O}_{1}]_{\a\b\c\d}^\sigma(0) 
 + \sum_i D^\sigma_i [\hat{\bar\mathcal{O}}_{i}]_{\a\b\c\d}^\sigma(0)  
 \Bigr\}
 \mathcal{J}^b_0(y) 
 \Bigr \rangle 
 \;. \la{C1xpt_su4}
\ee
The $\rmO(p^4)$ weak operators $\hat{\bar\mathcal{O}}_{i}$ here 
have the same chiral structures as the 27-plets of Ref.~\cite{p4},
listed in \eqs\nr{o27_2}--\nr{o27_25} of Appendix~B,  
but each of them comes in two variants after the valence 
flavour projection, corresponding to $\sigma=\pm$. 
The result can be written in the form 
\be
  {[\hat \mathcal{C}_{1}]}^{ab,\sigma}_{\a\b\c\d}(x_0,y_0) = 
  \hat \Delta^{ab,\sigma}_{\a\b\c\d} \biggl\{ 
  g_1^{\sigma} \Bigl[  \mathcal{C}(x_0) \mathcal{C}(y_0) 
 +  \sigma \mathcal{D}_{27}(x_0,y_0)  \Bigr]
 +  \mathcal{E}^\sigma (x_0,y_0)
 \biggr\} 
 \;, \la{Cpm} 
\ee
where $\hat \Delta^{ab,\sigma}_{\a\b\c\d}
\equiv 2 S^\sigma_{\a\b\c\d}(T^a,T^b)$, with the function 
$S^\sigma_{\a\b\c\d}(T^a,T^b)$ given in \eq\nr{Ssigma}.
The function $\mathcal{D}_{27}(x_0,y_0)$ 
is identical to the one for the 27-plet in \eq\nr{calD}.

The functions $\mathcal{E}^{\sigma}(x_0,y_0)$  in \eq\nr{Cpm}
contain the contributions of the $\rmO(p^4)$ low-energy constants, 
beyond those already contained in the factorized 
term $\mathcal{C}(x_0) \mathcal{C}(y_0)$:
\ba
 {\mathcal E}^\sigma(x_0,y_0) &=& 4 M^4 P'(x_0) P'(y_0) 
 \left[D^\sigma_2  +2 D^\sigma_{19} 
 - 4 D^\sigma_{24} - 4 D^\sigma_{25} \right] 
 + \nonumber\\
 &+& 4  M^6 P(x_0) P(y_0)
 \left[D^\sigma_4  + {\Nf\over 2} D^\sigma_{7} - D^\sigma_{20} - 
 {D^\sigma_{21}} - F^2 (\Nf L_4 +L_5) \right] 
 \;. \hspace*{1cm}
 \label{eq:p4_su4}
\ea
Taking into account that $\mathcal{C}(x_0)$ is finite, 
that the $\mathcal{D}_{27}$ contains the divergences
in \eq\nr{divD27}, 
that the QCD $\rmO(p^4)$ constants contain the divergence
in \eq\nr{divL4}, 
and that the weak $\rmO(p^4)$ constants contain the divergences
\ba
 D_4^\sigma &=& {D_4^{\sigma r}} + 
 g_1^{\sigma} \, F^2\lambda 
 \Bigl( 
 {\Nf+ 3\sigma \over 8} 
 \Bigr)
 \;, \\
 D_7^\sigma &=& {D_7^{\sigma r}} + 
 g_1^{\sigma} \, F^2\lambda 
 \Bigl( 
 {1 \over 4} 
 \Bigr)
 \;, \\
 D_{20}^\sigma &=& {D_{20}^{\sigma r}} +  
 g_1^{\sigma} \, F^2\lambda 
 \Bigl( 
 {\sigma \over 8} 
 \Bigr)
 \;, \\
 D_{24}^\sigma &=& {D_{24}^{\sigma r}} +  
 g_1^{\sigma} \, F^2\lambda 
 \Bigl( 
 {\sigma \over 32} 
 \Bigr)
 \;, \la{eq:duv_su4}
\ea
the correlation function $\hat\mathcal{C}_{1}$ in \eq\nr{Cpm}
can be seen to be finite.


\end{document}